\documentclass[aps,twocolumn,floats,prl,superscriptaddress,nofootinbib]{revtex4-1}

\usepackage{graphics,graphicx,epsfig}
\usepackage{amssymb,color}
\usepackage{epsf,epstopdf,wrapfig}
\usepackage{amsmath}
\usepackage{amsmath}
\usepackage{amssymb}
\usepackage{graphicx}
\usepackage{bm}
\usepackage{xr}
\usepackage{scrextend}
\definecolor{purple}{rgb}{0.58, 0.44, 0.86}
\usepackage{lipsum}

\newcommand{\beq}{\begin{equation}}
\newcommand{\eeq}{\end{equation}}
\newcommand{\bea}{\begin{eqnarray}}
\newcommand{\eea}{\end{eqnarray}}

\newcommand{\br}{ \mathbf{r}}
\newcommand{\bv}{ \mathbf{v}}
\newcommand{\bs}{ \mathbf{s}}

\newcommand{\bF}{ \mathbf{F}}

\newcommand{\bpi}{\boldsymbol{\pi}}
\newcommand{\bn}{\boldsymbol{n}}

\begin{document}

\title{Spin-Waves without Spin-Waves: A Case for Soliton Propagation in Starling Flocks}

\author{Andrea Cavagna}
\affiliation{Istituto Sistemi Complessi, Consiglio Nazionale delle Ricerche, UOS Sapienza, 00185 Rome, Italy}
\affiliation{Dipartimento di Fisica, Universit\`a\ Sapienza, 00185 Rome, Italy}
\affiliation{Istituto Nazionale di Fisica Nucleare, Sezione Roma 1, Rome, Italy}

\author{Guido Cimino}
\affiliation{Dipartimento di Fisica, Universit\`a\ Sapienza, 00185 Rome, Italy}
\affiliation{Istituto Sistemi Complessi, Consiglio Nazionale delle Ricerche, UOS Sapienza, 00185 Rome, Italy}

\author{Javier Crist\' in}
\affiliation{Departament de F\' isica, Universitat Aut\`onoma de Barcelona, 08193 Bellaterra, Spain}

\author{Matteo Fiorini}
\affiliation{Dipartimento di Fisica, Universit\`a\ Sapienza, 00185 Rome, Italy}
\affiliation{Istituto Sistemi Complessi, Consiglio Nazionale delle Ricerche, UOS Sapienza, 00185 Rome, Italy}

\author{Irene Giardina}
\affiliation{Dipartimento di Fisica, Universit\`a\ Sapienza, 00185 Rome, Italy}
\affiliation{Istituto Sistemi Complessi, Consiglio Nazionale delle Ricerche, UOS Sapienza, 00185 Rome, Italy}
\affiliation{Istituto Nazionale di Fisica Nucleare, Sezione Roma 1, Rome, Italy}

\author{Angelo Giustiniani}
\affiliation{Dipartimento di Fisica, Universit\`a\ Sapienza, 00185 Rome, Italy}
\affiliation{Istituto Sistemi Complessi, Consiglio Nazionale delle Ricerche, UOS Sapienza, 00185 Rome, Italy}

\author{Tom\'as S. Grigera}
\affiliation{Instituto de F\'\i{}sica de L\'\i{}quidos y Sistemas Biol\'ogicos CONICET -  Universidad Nacional de La Plata,  La Plata, Argentina}
\affiliation{CCT CONICET La Plata, Consejo Nacional de Investigaciones Cient\'\i{}ficas y T\'ecnicas, Argentina}
\affiliation{Departamento de F\'\i{}sica, Facultad de Ciencias Exactas, Universidad Nacional de La Plata, Argentina}
\affiliation{Istituto Sistemi Complessi, Consiglio Nazionale delle Ricerche, UOS Sapienza, 00185 Rome, Italy}
    
\author{Stefania Melillo}
\affiliation{Istituto Sistemi Complessi, Consiglio Nazionale delle Ricerche, UOS Sapienza, 00185 Rome, Italy}
\affiliation{Dipartimento di Fisica, Universit\`a\ Sapienza, 00185 Rome, Italy}

\author{Roberto A. Palombella}
\affiliation{Dipartimento di Fisica, Universit\`a\ Sapienza, 00185 Rome, Italy}
\affiliation{Istituto Sistemi Complessi, Consiglio Nazionale delle Ricerche, UOS Sapienza, 00185 Rome, Italy}

\author{Leonardo Parisi}
\affiliation{Istituto Sistemi Complessi, Consiglio Nazionale delle Ricerche, UOS Sapienza, 00185 Rome, Italy}
\affiliation{Dipartimento di Fisica, Universit\`a\ Sapienza, 00185 Rome, Italy}

\author{Antonio Ponno}
\affiliation{Dipartimento di Matematica T. Levi-Civita, Universit\`a degli Studi di Padova, 35131 Padova, Italy}

\author{Mattia Scandolo}
\affiliation{James Franck Institute, University of Chicago, 60637 Chicago, USA}

\author{Zachary S. Stamler}
\affiliation{Dipartimento di Fisica, Universit\`a\ Sapienza, 00185 Rome, Italy}
\affiliation{Istituto Sistemi Complessi, Consiglio Nazionale delle Ricerche, UOS Sapienza, 00185 Rome, Italy}

\begin{abstract}
Collective turns in starling flocks propagate linearly with negligible attenuation, indicating the existence of an underdamped sector in the dispersion relation. Beside granting linear propagation of the phase perturbations, the real part of the frequency should also yield a spin-wave form of the unperturbed correlation function.  However, new high-resolution experiments on real flocks show that underdamped traveling waves coexist with an overdamped Lorentzian correlation. 
Theory and experiments are reconciled once we add to the dynamics a Fermi-Pasta-Ulam-Tsingou term. 
\end{abstract}

\maketitle

Experiments on starling flocks \cite{attanasi+al_14} show that collective turns propagate across the group as linear waves with very little damping  (Videos S1-S2) and with a speed of propagation that is higher the larger the polarization; these  waves are not accompanied by significant density fluctuations, as the interaction network remains nearly unchanged during the turn. Moreover, birds turn along equal-radius  -- rather than parallel -- trajectories, which allows them to keep their speed within a physiological range \cite{pomeroy1992structure}. 
To explain this phenomenology it was introduced in  \cite{attanasi+al_14, cavagna+al_15} the Inertial Spin Model (ISM): at equilibrium the ISM belongs to Model G universality class \cite{hohenberg1977theory}, hence in its low temperature phase it sustains phase waves even in absence of density fluctuations (second sound). In the ISM the velocity $\bv_i$ is turned by the spin $\bs_{i}$, while the spin is acted upon by the alignment force,  
\bea 
\dot\bv_i &&= \frac{\partial H}{\partial \bs_i}  \times \bv_{i}
    \nonumber
     \ ,  \\
   \dot\bs_i && = - \bv_i \times \frac{\partial H}{\partial \bv_i} - \frac{\eta}{\chi} \bs_i +  \boldsymbol \xi_i \ ,
\label{ism}
\eea
where the pseudo-Hamiltonian is given by,
\beq
H = \sum_i \frac{\bs_i^2}{2\chi} + \frac{J}{4v_0^2} \sum_{ij} n_{ij} (\bv_i-\bv_j)^2 \ .
\eeq
The adjacency matrix $n_{ij}$ depends on time due to the self-propulsion equation, $\br_i = \bv_i$ (whence `pseudo'); the noise has correlator, $\langle \xi_i^{\mu}(t) \xi_j^{\nu}(t') \rangle = 2 \eta T \,\delta(t-t')\,\delta_{ij}\delta^{\mu\nu}$, and the velocity modulus is fixed, $|\bv_i| = v_0$. The parameter $J$ is the alignment coupling constant (or stiffness), while $\chi$ is the (generalized) inertia and $\eta$ is the nonconservative friction.
The ISM's overdamped limit is the Vicsek model \cite{vicsek+al_95}.
In the low-temperature symmetry-broken phase the velocities predominantly point in the same mean direction, $\bv=(1/N) \sum_i \bv_i$, and we can expand in the small transverse fluctuations, $\bpi_i$, writing, 
$\bv_i=\bv_i^L + \boldsymbol{\pi}_i$, 
where 
$|\boldsymbol{\pi}_i | \ll |\bv|$ 
and  
$| \bv_i^L | = \sqrt{v_0^2-\bpi_i^2}$, 
with 
$\boldsymbol{\pi}_i \cdot \bv=0$
\cite{dyson_56}. 
By expanding linearly in $\bpi_i$ the homogeneous equation of motion for the velocity (see Supplemental Material -- SM) we obtain,
\begin{equation}
   \chi \,\ddot\bpi_i + \eta\,\dot\bpi_i  + J\sum_j \Lambda_{ij} \bpi_j = 0 \ ,
 \label{eqpi}
\end{equation}
where $\Lambda_{ij}=-n_{ij}+\delta_{ij}\sum_kn_{ik}$ is the discrete Laplacian. By assuming that $n_{ij}(t)$ varies slowly compared to the velocity's relaxation time \cite{mora2016local} and by going to the continuum limit, $\Lambda_{ij}\to - a^2\nabla^2$ (where $a$ is the mean interparticle distance), we can derive the quasi-equilibrium dispersion relation of the ISM,
\begin{equation}
{\omega}(k)=-i\gamma \pm \sqrt{(J/\chi) a^2 k^2 - \gamma^2 } \ ,
\label{brasilia}
\end{equation}
where $\gamma = \eta/2\chi$ is the reduced friction.
In the underdamped regime the frequency develops a real part, which for $(J/\chi) a^2 k^2 \gg \gamma^2$ becomes $\omega(k) \approx \pm c_s k$, with $c_s = a\,\sqrt{J/\chi}$; hence, the system sustains linearly propagating waves with speed $c_s$. Simulations confirm that within this deeply underdamped phase the change of direction of one particle elicits a collective turn of the entire group, while in the overdamped phase, $(J/\chi) a^2 k^2 < \gamma^2$, damping and sub-linear diffusion cause disgregation of the flock, with no collective turning; these results are evident from the acceleration curves of the particles (Fig.1) and  even more vividly from the corresponding Videos S3-S4. The speed of propagation in real flocks was found to be in line with the ISM prediction: in the ordered phase the polarization $\Phi = |\bv|/v_0$, is connected to the transverse fluctuations by the relation, $(1- \Phi) \approx \langle \bpi^2/v_0^2\rangle/2 \sim T/J$ (the prefactors depend on the specifics of the lattice \cite{dyson_56, bialek+al_12, cavagna2016spatio}), whence we obtain $c_s \sim 1/\sqrt{1-\Phi}$, a prediction in fair agreement with experiments on starling flocks \cite{attanasi+al_14}.

%%%%%%%%%%%%%%%%%%%%%%%%%%%%%%%%%%%%%%%%%%
\begin{figure}
\centering
\includegraphics[width=0.5 \textwidth]{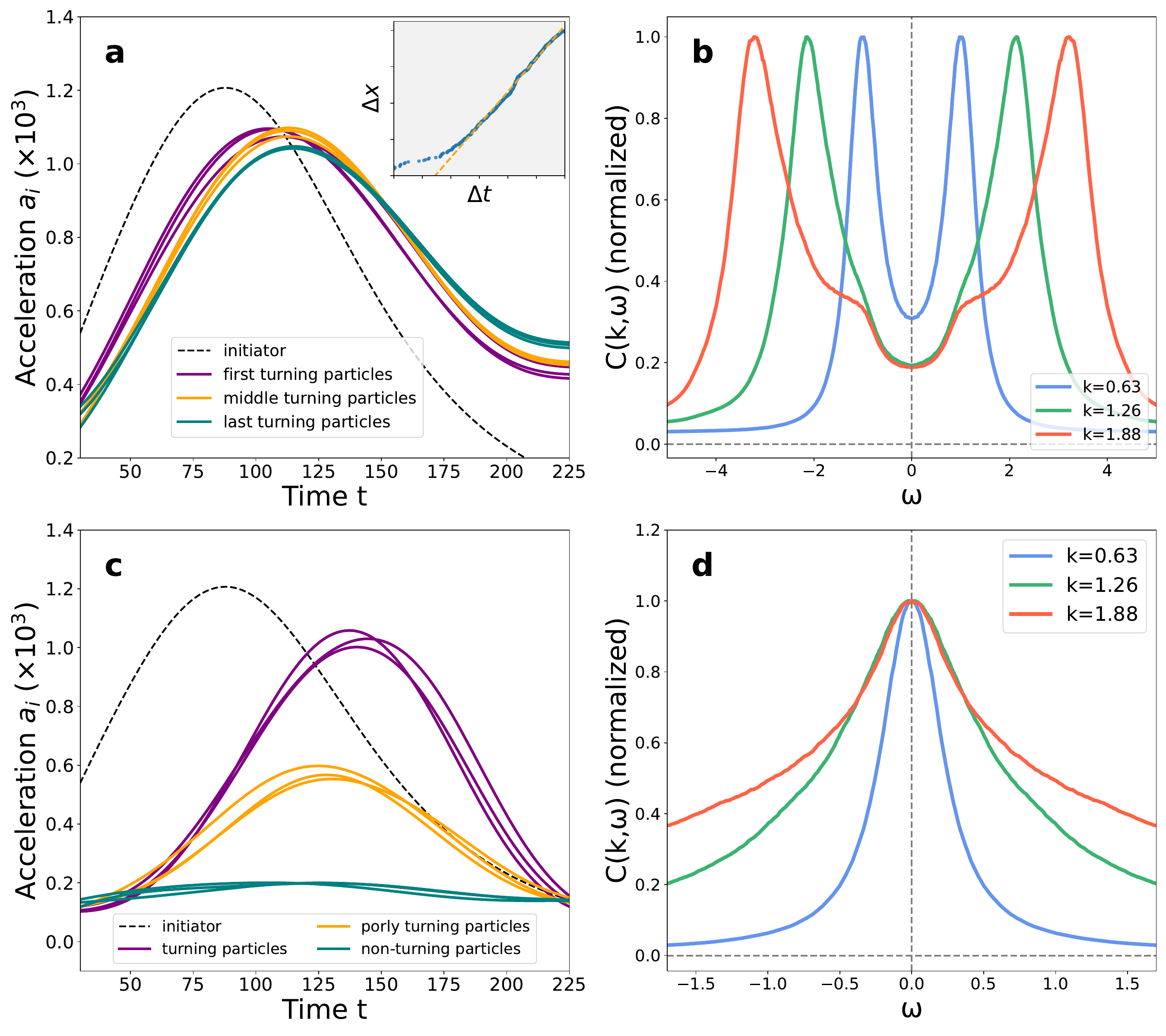}
\caption{{\bf Simulations - ISM.} 
Upper panels: underdamped regime. 
a) The initiator is turned by $\pi/2$, eliciting a collective turn: the accelerations of all the particles show coherent turning, with little damping (Video S3). Inset: traveled distance of the perturbation (from an intercept on the phase) vs time indicates linear propagation. 
b) The velocity correlation function has clear spin-wave peaks at $\omega_{SW}= \pm c_s k$ (SM).
Lower panels: overdamped regime. 
c) The initiator fails to elicit a collective turn; the accelerations of the other particles crash due to damping (Video S4). 
d) The velocity correlation function is Lorentzian, with half-width frequency $\omega_{L} \sim k^2$ (SM). 
Simulations are performed in $d=3$.
}
\label{simu}
\end{figure}
%%%%%%%%%%%%%%%%%%%%%%%%%%%%%%%%%%%%%%%%%%%

The existence of a real part of the frequency in the underdamped regime has another consequence, namely that the correlation function of the velocity fluctuations develops two spin-wave (SW) peaks at $\omega_\mathrm{sw} = \pm c_s k$; conversely, in the overdamped phase, $c_s^2\,  k^2 < \gamma^2$, the correlation function is quasi-Lorentzian (L), with half-width frequency, $\omega_\mathrm{L} =(J/\eta)a^2k^2$. In the Supplemental Material we perform a linear analysis of the out-of-equilibrium active field theory corresponding to \eqref{ism}, in which $n_{ij}(t)$ is {\it not} assumed to be constant, hence including density fluctuations, and show that also the active case conforms to the standard spin-wave phenomenology. Simulations of the ISM confirm this scenario (Fig.1). Note that the association of spin-wave peaks and traveling waves does not depend on the specific way propagating modes emerge, whether it is through inertia, as in the ISM, or through the  velocity-density coupling giving rise to sound modes as in the Vicsek model  \cite{toner+al_95, tu1998sound, toner1998flocks}: whenever the frequency develops a real part, there should be both spin-wave/sound propagation {\it and} spin-wave/sound peaks; this is the core of linear response theory.

Experimental evidence on starling flocks starkly disagrees with this scenario, though. To calculate the correlation function accurately enough to resolve the potential spin-wave peaks one needs a small frequency step, which requires experimental acquisitions with time duration significantly longer than \cite{attanasi+al_14}; to this end we developed a new panning 3D stereo system and run new experiments (Fig.2a-d and SM). Surprisingly enough, the correlation functions computed from this new dataset (an example in Fig.2f, more in SM) do not show any trace of spin-wave peaks; instead, they have a very much Lorentzian central peak at $\omega=0$, with half-width frequency $\omega_\mathrm{L}\sim k^2$ (SM), both clear hallmarks of an overdamped system. And yet, phase perturbations {\it do} propagate linearly across the group, eliciting collective turns (Fig.2e). So, starling flocks develop traveling spin-waves under perturbation, without spin-waves in the spontaneous fluctuations. How is that possible?

%%%%%%%%%%%%%%%%%%%%%%%%%%%%%%%%%%%%%%%%%%
\begin{figure}
\centering
\includegraphics[width=0.5 \textwidth]{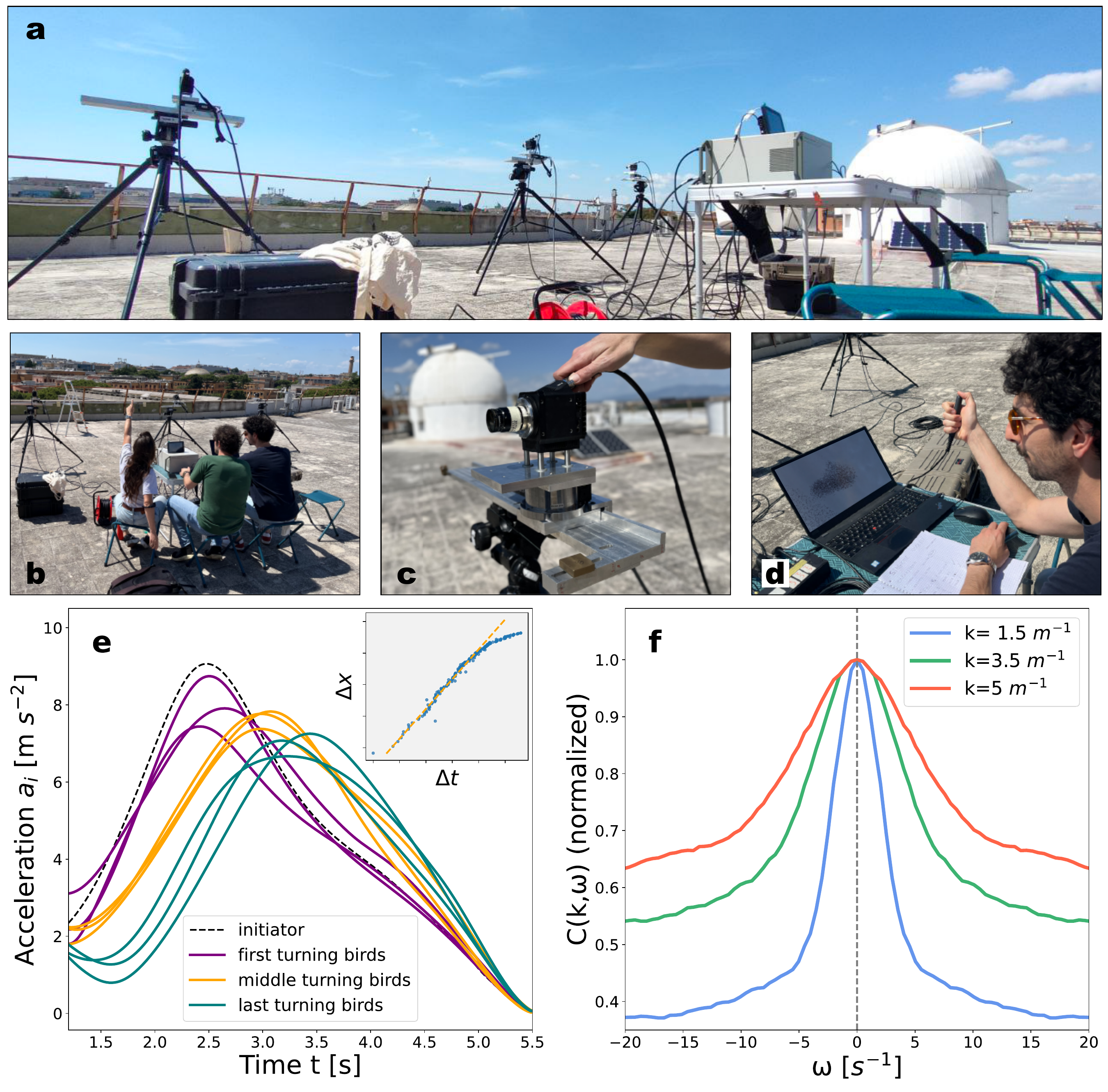}
\caption{{\bf Experiments - Starling flocks.} 
a-d) In the new experimental setup micrometric rotors (c) are used to pan the stereo camera system and track flocks for longer time intervals (up to 12.5 sec vs 3 sec of the old experiments -- SM).
e) In turning flocks phase perturbations propagate linearly with negligible dissipation, as in the underdamped phase of the ISM (Fig.1a), and yet: 
f) The correlation function has a quasi-Lorentzian form, with no spin-wave peaks, and an half-width frequency scaling as $\omega_\mathrm{L}\sim k^2$ (see SM), as in the overdamped phase of the ISM (Fig.1d).
}
\label{expe}
\end{figure}
%%%%%%%%%%%%%%%%%%%%%%%%%%%%%%%%%%%%%%%%%%%

There are two possible ways out of this apparent paradox: either some further out-of-equilibrium effect radically changes the spin-wave prediction, or the relationship between response to a perturbation and unperturbed correlation is not the one we expected. The first hypothesis is unlikely, as the inclusion of out-of-equilibrium density fluctuations at the linear level does not change the classic spin-wave scenario (SM) and simulations thoroughly confirm that. The second possibility seems absurd as long as we stay within the boundaries of linear response theory. But, if a strong nonlinearity is present, then things could change -- even at equilibrium.

The strong polarization of bird flocks is due to the fact that spontaneous fluctuations are weak; on the other hand, a localized phase fluctuation due to a perturbation -- and in particular  one able to elicit a collective turn -- may be much stronger: flocks typically turn because some individual perceives an external threat, hence changing heading in a robust way compared to its normal cruising fluctuations. A possible way to make progress, then, is to hypothesize that individuals respond weakly to small fluctuations in the orientation of their neighbours -- which are perceived as normal business -- and very strongly to large fluctuations, which may signal a threat. This would imply that the stiffness, i.e. the alignment coupling constant, is not in fact a constant, but it has a component that increases when the local phase distortion, $\delta\varphi$, increases. 
In this way, due to their small stiffness, spontaneous fluctuations would be overdamped, showing no spin-wave peaks  in the correlation function, while when a perturbation strikes the local phase distortion jumps, the stiffness suddenly increases and the system becomes underdamped, therefore able to propagate the phase wave necessary for the flock to collectively turn.

We choose the simplest way to implement this mechanism, namely to introduce a non-Gaussian term in the $\bv$-dependent part of the Hamiltonian,
\beq
H_{\bv} = \frac{J}{4v_0^2} \sum_{ij} n_{ij} (\bv_i-\bv_j)^2 + \frac{J_4}{4v_0^4}\,  \sum_{ij} n_{ij} (\bv_i-\bv_j)^4 \ .
\label{pepito}
\eeq
For the sake of the following arguments it is convenient to consider planar velocities, hence transverse fluctuations $\bpi_i$ become scalars and we can 
expand \eqref{pepito} in the dimensionless phases, $\varphi_i = \pi_i/v_0$, to obtain (see  SM),
\beq
H_\bv \approx  \frac14 \sum_{ij} n_{ij}\, (\varphi_i-\varphi_j)^2 \left[J +  J_4 (\varphi_i-\varphi_j)^2 \right] \ ,
\label{todi}
\eeq
which naturally invites us to define an effective stiffness,
\beq
J_\mathrm{eff} = J + J_4 \; \delta\varphi^2  \ .
\label{jeff}
\eeq
Relation \eqref{jeff} aligns with our  purpose of having a stiffness that sharply increases with the local phase distortion. But can we  find a 
sector in the parameters space in which this mechanism may actually work?

In the unperturbed regime we want spontaneous fluctuations to be overdamped; we leave this job to the standard stiffness, hence requiring $J_\mathrm{eff} \sim J$ and  $(J/\chi)\, a^2k^2 \lesssim \gamma^2$ (near-critical damping would still produce a quasi-Lorentzian correlation function, hence we do not need to asymptotically push this inequality); by using $k\sim 1/L$ as a relevant scale, we get overdamped correlation for $J \lesssim \chi\,\gamma^2  (L/a)^2$; moreover, to achieve $J_\mathrm{eff} \sim J$ we require $J_4\, \delta\varphi^2 \lesssim J$; given that $\langle\delta\varphi^2\rangle \sim (1- \Phi)$, we need $J_4 \lesssim J/(1-\Phi)$. 
On the other hand, when a strong perturbation strikes, producing a sizeable localized phase change, $\delta\varphi \sim O(1)$, we want the quartic term to kick in, namely $J_\mathrm{eff} \sim J_4$, which requires $J_4 \gg J$. Once the new term has been activated, we need its dynamics to be underdamped to grant propagation of the phase distortion; we do not know the dispersion relation in the fully non-linear regime, but we can (very crudely) substitute $J$ with $J_\mathrm{eff}\sim J_4$ in \eqref{brasilia}, thus obtaining $J_4 \gg  \chi\,\gamma^2  (L/a)^2$ (to have linear propagation we {\it do} need this inequality to be asymptotic).
We can rearrange these conditions as, 
\beq
J \; \lesssim \; \chi\,\gamma^2  (L/a)^2\;  \ll \; J_4 \; \lesssim \; J/(1-\Phi)  \ .
\label{gnonto}
\eeq
Simply stated, $J_4$ must be large enough to rule the transport of finite phase perturbations, but small enough to remain dormant under spontaneous phase fluctuations.
The fact that $1/(1-\Phi) \gg 1$ leaves us some margin to satisfy \eqref{gnonto}, suggesting that \eqref{pepito} may actually yield a non-linear regime of ``spin-waves without spin-waves''.

%%%%%%%%%%%%%%%%%%%%%%%%%%%%%%%%%%%%%%%%%%
\begin{figure}
\centering
\includegraphics[width=0.5 \textwidth]{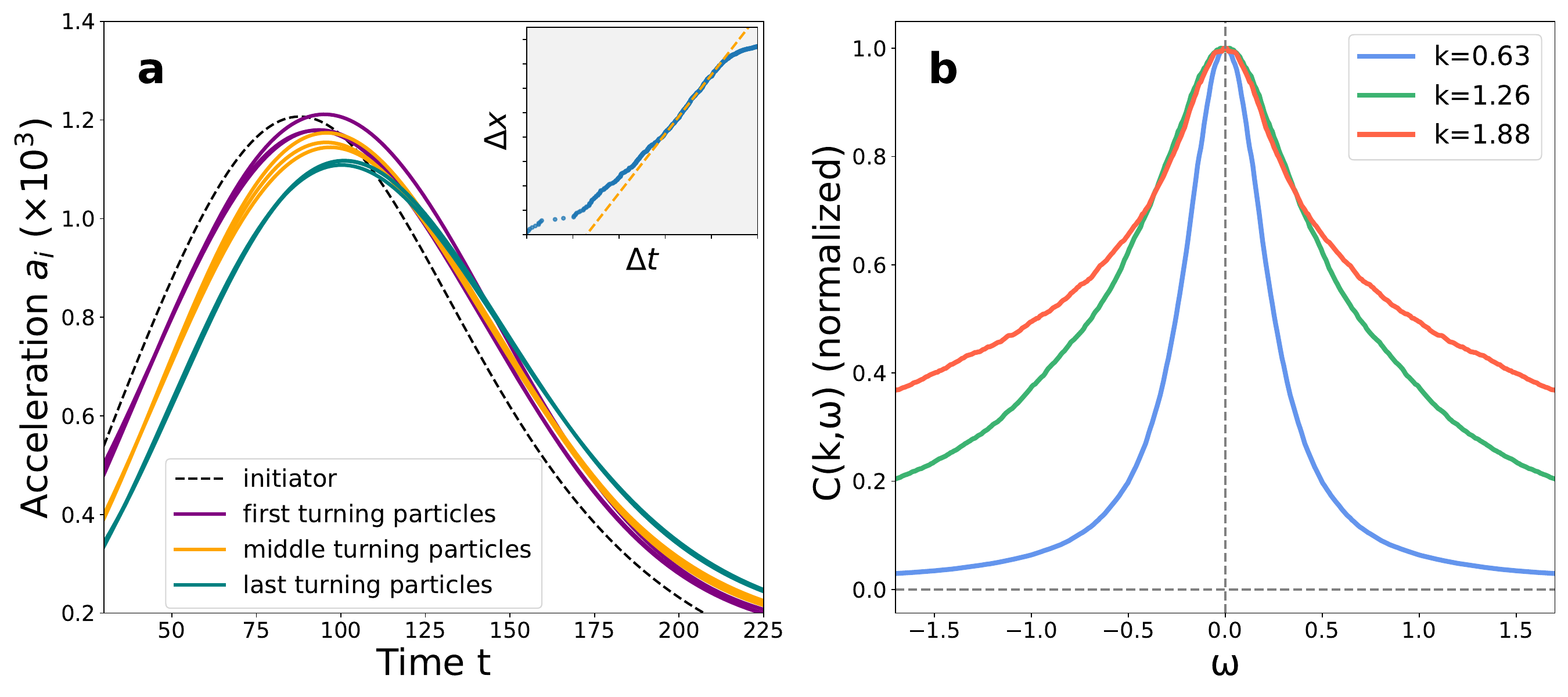}
\caption{{\bf Simulations - ISM + FPUT.} 
a) When a quartic term is added to the interaction, linear underdamped propagation of phase perturbations is achieved (Video S5), and:
b) Spontaneous fluctuations remain overdamped, as in natural flocks (Fig.2). Simulations are performed in $d=3$.
}
\label{FPU}
\end{figure}
%%%%%%%%%%%%%%%%%%%%%%%%%%%%%%%%%%%%%%%%%%%

Numerical simulations of \eqref{ism}-\eqref{pepito} in the general $3d$ non-planar case confirm this theoretical expectation (Fig.3). When the polarization is high, the quadratic stiffness $J$ low and the nonlinear coupling $J_4$ large, we obtain the same phenomenology as in real experiments: in absence of external perturbations we find a Lorentzian correlation function, with half-width frequency $\omega_\mathrm{L} \sim k^2$ (SM); at the same time, when one individual turns, there is linear wave propagation, eliciting a collective turn of the entire group (Fig.3 and Video S5). 
Spontaneous fluctuations, ruled by the low stiffness $J$, are overdamped, while external perturbations respond to the large nonlinear stiffness $J_4$, which makes them underdamped. 
Moreover, the relation $(1- \Phi) \approx\langle\bpi^2/v_0^2\rangle/2$, depends only on the expansion for small $\bpi_i$, hence if we extend to $J_\mathrm{eff}$ the standard low-$T$ relations, $\langle\bpi^2/v_0^2\rangle \sim T/J_\mathrm{eff}$ and $c_s \sim\sqrt{J_\mathrm{eff}}$, we obtain the same prediction as in the standard ISM for how the speed of propagation should depend on the polarization, $c_s\sim 1/\sqrt{1-\Phi}$, which -- as we already remarked -- is well verified by experiments on real flocks \cite{attanasi+al_14}.

It is interesting to note that the new quartic interaction in \eqref{pepito} is very similar in spirit to the quartic speed control mechanism added to the Hamiltonian in 
\cite{cavagna2019low, cavagna2022marginal, cavagna2022renormalization} to explain scale-free correlations of speed fluctuations in flocks, 
$H_\mathrm{speed\,ctrl.} \approx \lambda \sum_i (v_i - v_0)^4$, where $v_0$ is the re\-fe\-rence physiological value of the speed.
The fact that birds seem to use this mechanism both to regulate their level of response to inter-particle misalignments and to control the deviations of their individual speed with respect to a physiological baseline, suggests that such a mechanism may embody a natural tendency of individuals  within biological groups to respond nonlinearly to external stimuli, by almost ignoring small fluctuations, while sharply reacting to large perturbations.

When we explicitly write down the equations of motion for the planar phases $\varphi_i$ in the new model (see SM), and neglect both noise and dissipation, we obtain evolution equations identical in form to those of the Fermi-Pasta-Ulam-Tsingou (FPUT) model \cite{fermi1955studies, gallavotti2008fermi}, 
\begin{equation}
   \ddot{{\varphi}}_i +  \sum_j n_{ij} (\varphi_i-\varphi_j)  +
2\frac{J_4}{J} \sum_j n_{ij} (\varphi_i-\varphi_j)^3 = 0  \ .
\label{FPUT}
\end{equation}
These equations are, up to a time rescaling $t\to \sqrt{J/\chi}\ t$, the Hamilton equations associated to the Hamiltonian
$H(\varphi,s)=\frac{1}{2\chi}\sum_i s_i^2+H_\bv$, where the `potential' energy $H_\bv$ is given by \eqref{todi}.
A well-known property of the FPUT dynamics is that the competition between nonlinearity and dispersion, ruled by the ratio $J_4/J$, allows for the propagation of solitary waves, or {\it solitons}, in both one and two dimensional lattices \cite{friesecke1994existence, Friesecke2003geometric, Chen2018Discrete}.
Although the problem at hand requires to study FPUT solitons on an adiabatically evolving $3d$ interaction network -- to take into account activity -- and to consider the theoretical effects of weak noise \cite{ORLOWSKI1989solitons} and dissipation \cite{Purwins2010dissipative} on the general FPUT mechanism, it is very tempting to interpret the traveling waves that we obtain in the new model and that we observe in natural flocks as FPUT solitons.

In conclusion, we have conducted new experiments on starlings flocks, finding a puzzling phenomenology: finite disturbances of the phase are linearly transported, giving rise to collective turns, while spontaneous phase fluctuations are overdamped, producing a Lorentzian correlation function with no evidence of the standard spin-wave peaks that normally accompany traveling waves. The ISM, originally introduced to account for underdamped linear propagation, cannot explain an overdamped correlation function. To reconcile experiments with theory we have added a quartic term to the alignment interaction, thanks to which weak spontaneous fluctuations of the phase are overdamped, while finite phase perturbations -- caused by external stimuli -- are underdamped, and therefore propagate linearly. The power unit of the new dynamics is a FPUT  core, suggesting that directional information within natural flocks of starlings is transported by nonlinear solitary waves.

We thank William Bialek, Enzo Branchini, Fabio Cecconi, Giulio Costantini and Mario Veca for insightful discussions. This work was supported by ERC Grant RG.BIO (Contract n. 785932), 
MIUR Grant INFO.BIO (Protocol n. R18JNYYMEY), and MIUR Grant PRIN2020 (Contract n. 2020PFCXPE-005).

\bibliographystyle{apsrev4-1}
\bibliography{general_cobbs_bibliography_file}

\begin{thebibliography}{38}%
\makeatletter
\providecommand \@ifxundefined [1]{%
 \@ifx{#1\undefined}
}%
\providecommand \@ifnum [1]{%
 \ifnum #1\expandafter \@firstoftwo
 \else \expandafter \@secondoftwo
 \fi
}%
\providecommand \@ifx [1]{%
 \ifx #1\expandafter \@firstoftwo
 \else \expandafter \@secondoftwo
 \fi
}%
\providecommand \natexlab [1]{#1}%
\providecommand \enquote  [1]{``#1''}%
\providecommand \bibnamefont  [1]{#1}%
\providecommand \bibfnamefont [1]{#1}%
\providecommand \citenamefont [1]{#1}%
\providecommand \href@noop [0]{\@secondoftwo}%
\providecommand \href [0]{\begingroup \@sanitize@url \@href}%
\providecommand \@href[1]{\@@startlink{#1}\@@href}%
\providecommand \@@href[1]{\endgroup#1\@@endlink}%
\providecommand \@sanitize@url [0]{\catcode `\\12\catcode `\$12\catcode
  `\&12\catcode `\#12\catcode `\^12\catcode `\_12\catcode `\%12\relax}%
\providecommand \@@startlink[1]{}%
\providecommand \@@endlink[0]{}%
\providecommand \url  [0]{\begingroup\@sanitize@url \@url }%
\providecommand \@url [1]{\endgroup\@href {#1}{\urlprefix }}%
\providecommand \urlprefix  [0]{URL }%
\providecommand \Eprint [0]{\href }%
\providecommand \doibase [0]{http://dx.doi.org/}%
\providecommand \selectlanguage [0]{\@gobble}%
\providecommand \bibinfo  [0]{\@secondoftwo}%
\providecommand \bibfield  [0]{\@secondoftwo}%
\providecommand \translation [1]{[#1]}%
\providecommand \BibitemOpen [0]{}%
\providecommand \bibitemStop [0]{}%
\providecommand \bibitemNoStop [0]{.\EOS\space}%
\providecommand \EOS [0]{\spacefactor3000\relax}%
\providecommand \BibitemShut  [1]{\csname bibitem#1\endcsname}%
\let\auto@bib@innerbib\@empty
%</preamble>
\bibitem [{\citenamefont {Attanasi}\ \emph {et~al.}(2014)\citenamefont
  {Attanasi}, \citenamefont {Cavagna}, \citenamefont {Del~Castello},
  \citenamefont {Giardina}, \citenamefont {Grigera}, \citenamefont {Jeli{\'c}},
  \citenamefont {Melillo}, \citenamefont {Parisi}, \citenamefont {Pohl},
  \citenamefont {Shen} \emph {et~al.}}]{attanasi+al_14}%
  \BibitemOpen
  \bibfield  {author} {\bibinfo {author} {\bibfnamefont {A.}~\bibnamefont
  {Attanasi}}, \bibinfo {author} {\bibfnamefont {A.}~\bibnamefont {Cavagna}},
  \bibinfo {author} {\bibfnamefont {L.}~\bibnamefont {Del~Castello}}, \bibinfo
  {author} {\bibfnamefont {I.}~\bibnamefont {Giardina}}, \bibinfo {author}
  {\bibfnamefont {T.~S.}\ \bibnamefont {Grigera}}, \bibinfo {author}
  {\bibfnamefont {A.}~\bibnamefont {Jeli{\'c}}}, \bibinfo {author}
  {\bibfnamefont {S.}~\bibnamefont {Melillo}}, \bibinfo {author} {\bibfnamefont
  {L.}~\bibnamefont {Parisi}}, \bibinfo {author} {\bibfnamefont
  {O.}~\bibnamefont {Pohl}}, \bibinfo {author} {\bibfnamefont {E.}~\bibnamefont
  {Shen}},  \emph {et~al.},\ }\href@noop {} {\bibfield  {journal} {\bibinfo
  {journal} {Nature Physics}\ }\textbf {\bibinfo {volume} {10}},\ \bibinfo
  {pages} {691} (\bibinfo {year} {2014})}\BibitemShut {NoStop}%
\bibitem [{\citenamefont {Pomeroy}\ and\ \citenamefont
  {Heppner}(1992)}]{pomeroy1992structure}%
  \BibitemOpen
  \bibfield  {author} {\bibinfo {author} {\bibfnamefont {H.}~\bibnamefont
  {Pomeroy}}\ and\ \bibinfo {author} {\bibfnamefont {F.}~\bibnamefont
  {Heppner}},\ }\href@noop {} {\bibfield  {journal} {\bibinfo  {journal} {The
  Auk}\ ,\ \bibinfo {pages} {256}} (\bibinfo {year} {1992})}\BibitemShut
  {NoStop}%
\bibitem [{\citenamefont {Cavagna}\ \emph
  {et~al.}(2015{\natexlab{a}})\citenamefont {Cavagna}, \citenamefont
  {Del~Castello}, \citenamefont {Giardina}, \citenamefont {Grigera},
  \citenamefont {Jelic}, \citenamefont {Melillo}, \citenamefont {Mora},
  \citenamefont {Parisi}, \citenamefont {Silvestri}, \citenamefont {Viale}
  \emph {et~al.}}]{cavagna+al_15}%
  \BibitemOpen
  \bibfield  {author} {\bibinfo {author} {\bibfnamefont {A.}~\bibnamefont
  {Cavagna}}, \bibinfo {author} {\bibfnamefont {L.}~\bibnamefont
  {Del~Castello}}, \bibinfo {author} {\bibfnamefont {I.}~\bibnamefont
  {Giardina}}, \bibinfo {author} {\bibfnamefont {T.}~\bibnamefont {Grigera}},
  \bibinfo {author} {\bibfnamefont {A.}~\bibnamefont {Jelic}}, \bibinfo
  {author} {\bibfnamefont {S.}~\bibnamefont {Melillo}}, \bibinfo {author}
  {\bibfnamefont {T.}~\bibnamefont {Mora}}, \bibinfo {author} {\bibfnamefont
  {L.}~\bibnamefont {Parisi}}, \bibinfo {author} {\bibfnamefont
  {E.}~\bibnamefont {Silvestri}}, \bibinfo {author} {\bibfnamefont
  {M.}~\bibnamefont {Viale}},  \emph {et~al.},\ }\href@noop {} {\bibfield
  {journal} {\bibinfo  {journal} {Journal of Statistical Physics}\ }\textbf
  {\bibinfo {volume} {158}},\ \bibinfo {pages} {601} (\bibinfo {year}
  {2015}{\natexlab{a}})}\BibitemShut {NoStop}%
\bibitem [{\citenamefont {Hohenberg}\ and\ \citenamefont
  {Halperin}(1977)}]{hohenberg1977theory}%
  \BibitemOpen
  \bibfield  {author} {\bibinfo {author} {\bibfnamefont {P.~C.}\ \bibnamefont
  {Hohenberg}}\ and\ \bibinfo {author} {\bibfnamefont {B.~I.}\ \bibnamefont
  {Halperin}},\ }\href@noop {} {\bibfield  {journal} {\bibinfo  {journal}
  {Reviews of Modern Physics}\ }\textbf {\bibinfo {volume} {49}},\ \bibinfo
  {pages} {435} (\bibinfo {year} {1977})}\BibitemShut {NoStop}%
\bibitem [{\citenamefont {Vicsek}\ \emph {et~al.}(1995)\citenamefont {Vicsek},
  \citenamefont {Czir{\'o}k}, \citenamefont {Ben-Jacob}, \citenamefont
  {Cohen},\ and\ \citenamefont {Shochet}}]{vicsek+al_95}%
  \BibitemOpen
  \bibfield  {author} {\bibinfo {author} {\bibfnamefont {T.}~\bibnamefont
  {Vicsek}}, \bibinfo {author} {\bibfnamefont {A.}~\bibnamefont {Czir{\'o}k}},
  \bibinfo {author} {\bibfnamefont {E.}~\bibnamefont {Ben-Jacob}}, \bibinfo
  {author} {\bibfnamefont {I.}~\bibnamefont {Cohen}}, \ and\ \bibinfo {author}
  {\bibfnamefont {O.}~\bibnamefont {Shochet}},\ }\href@noop {} {\bibfield
  {journal} {\bibinfo  {journal} {Phys Rev Lett}\ }\textbf {\bibinfo {volume}
  {75}},\ \bibinfo {pages} {1226} (\bibinfo {year} {1995})}\BibitemShut
  {NoStop}%
\bibitem [{\citenamefont {Dyson}(1956)}]{dyson_56}%
  \BibitemOpen
  \bibfield  {author} {\bibinfo {author} {\bibfnamefont {F.}~\bibnamefont
  {Dyson}},\ }\href@noop {} {\bibfield  {journal} {\bibinfo  {journal}
  {Physical review}\ }\textbf {\bibinfo {volume} {102}},\ \bibinfo {pages}
  {1217} (\bibinfo {year} {1956})}\BibitemShut {NoStop}%
\bibitem [{\citenamefont {Mora}\ \emph {et~al.}(2016)\citenamefont {Mora},
  \citenamefont {Walczak}, \citenamefont {Del~Castello}, \citenamefont
  {Ginelli}, \citenamefont {Melillo}, \citenamefont {Parisi}, \citenamefont
  {Viale}, \citenamefont {Cavagna},\ and\ \citenamefont
  {Giardina}}]{mora2016local}%
  \BibitemOpen
  \bibfield  {author} {\bibinfo {author} {\bibfnamefont {T.}~\bibnamefont
  {Mora}}, \bibinfo {author} {\bibfnamefont {A.~M.}\ \bibnamefont {Walczak}},
  \bibinfo {author} {\bibfnamefont {L.}~\bibnamefont {Del~Castello}}, \bibinfo
  {author} {\bibfnamefont {F.}~\bibnamefont {Ginelli}}, \bibinfo {author}
  {\bibfnamefont {S.}~\bibnamefont {Melillo}}, \bibinfo {author} {\bibfnamefont
  {L.}~\bibnamefont {Parisi}}, \bibinfo {author} {\bibfnamefont
  {M.}~\bibnamefont {Viale}}, \bibinfo {author} {\bibfnamefont
  {A.}~\bibnamefont {Cavagna}}, \ and\ \bibinfo {author} {\bibfnamefont
  {I.}~\bibnamefont {Giardina}},\ }\href@noop {} {\bibfield  {journal}
  {\bibinfo  {journal} {Nature Physics}\ }\textbf {\bibinfo {volume} {12}},\
  \bibinfo {pages} {1153} (\bibinfo {year} {2016})}\BibitemShut {NoStop}%
\bibitem [{\citenamefont {Bialek}\ \emph {et~al.}(2012)\citenamefont {Bialek},
  \citenamefont {Cavagna}, \citenamefont {Giardina}, \citenamefont {Mora},
  \citenamefont {Silvestri}, \citenamefont {Viale},\ and\ \citenamefont
  {Walczak}}]{bialek+al_12}%
  \BibitemOpen
  \bibfield  {author} {\bibinfo {author} {\bibfnamefont {W.}~\bibnamefont
  {Bialek}}, \bibinfo {author} {\bibfnamefont {A.}~\bibnamefont {Cavagna}},
  \bibinfo {author} {\bibfnamefont {I.}~\bibnamefont {Giardina}}, \bibinfo
  {author} {\bibfnamefont {T.}~\bibnamefont {Mora}}, \bibinfo {author}
  {\bibfnamefont {E.}~\bibnamefont {Silvestri}}, \bibinfo {author}
  {\bibfnamefont {M.}~\bibnamefont {Viale}}, \ and\ \bibinfo {author}
  {\bibfnamefont {A.~M.}\ \bibnamefont {Walczak}},\ }\href {\doibase
  10.1073/pnas.1118633109} {\bibfield  {journal} {\bibinfo  {journal} {Proc
  Natl Acad Sci USA}\ }\textbf {\bibinfo {volume} {109}},\ \bibinfo {pages}
  {4786} (\bibinfo {year} {2012})}\BibitemShut {NoStop}%
\bibitem [{\citenamefont {Cavagna}\ \emph {et~al.}(2016)\citenamefont
  {Cavagna}, \citenamefont {Conti}, \citenamefont {Giardina}, \citenamefont
  {Grigera}, \citenamefont {Melillo},\ and\ \citenamefont
  {Viale}}]{cavagna2016spatio}%
  \BibitemOpen
  \bibfield  {author} {\bibinfo {author} {\bibfnamefont {A.}~\bibnamefont
  {Cavagna}}, \bibinfo {author} {\bibfnamefont {D.}~\bibnamefont {Conti}},
  \bibinfo {author} {\bibfnamefont {I.}~\bibnamefont {Giardina}}, \bibinfo
  {author} {\bibfnamefont {T.~S.}\ \bibnamefont {Grigera}}, \bibinfo {author}
  {\bibfnamefont {S.}~\bibnamefont {Melillo}}, \ and\ \bibinfo {author}
  {\bibfnamefont {M.}~\bibnamefont {Viale}},\ }\href@noop {} {\bibfield
  {journal} {\bibinfo  {journal} {Physical Biology}\ }\textbf {\bibinfo
  {volume} {13}},\ \bibinfo {pages} {065001} (\bibinfo {year}
  {2016})}\BibitemShut {NoStop}%
\bibitem [{\citenamefont {Toner}\ and\ \citenamefont {Tu}(1995)}]{toner+al_95}%
  \BibitemOpen
  \bibfield  {author} {\bibinfo {author} {\bibfnamefont {J.}~\bibnamefont
  {Toner}}\ and\ \bibinfo {author} {\bibfnamefont {Y.}~\bibnamefont {Tu}},\
  }\href@noop {} {\bibfield  {journal} {\bibinfo  {journal} {Phys Rev Lett}\
  }\textbf {\bibinfo {volume} {75}},\ \bibinfo {pages} {4326} (\bibinfo {year}
  {1995})}\BibitemShut {NoStop}%
\bibitem [{\citenamefont {Tu}\ \emph {et~al.}(1998{\natexlab{a}})\citenamefont
  {Tu}, \citenamefont {Toner},\ and\ \citenamefont {Ulm}}]{tu1998sound}%
  \BibitemOpen
  \bibfield  {author} {\bibinfo {author} {\bibfnamefont {Y.}~\bibnamefont
  {Tu}}, \bibinfo {author} {\bibfnamefont {J.}~\bibnamefont {Toner}}, \ and\
  \bibinfo {author} {\bibfnamefont {M.}~\bibnamefont {Ulm}},\ }\href@noop {}
  {\bibfield  {journal} {\bibinfo  {journal} {Physical review letters}\
  }\textbf {\bibinfo {volume} {80}},\ \bibinfo {pages} {4819} (\bibinfo {year}
  {1998}{\natexlab{a}})}\BibitemShut {NoStop}%
\bibitem [{\citenamefont {Toner}\ and\ \citenamefont
  {Tu}(1998)}]{toner1998flocks}%
  \BibitemOpen
  \bibfield  {author} {\bibinfo {author} {\bibfnamefont {J.}~\bibnamefont
  {Toner}}\ and\ \bibinfo {author} {\bibfnamefont {Y.}~\bibnamefont {Tu}},\
  }\href@noop {} {\bibfield  {journal} {\bibinfo  {journal} {Physical review
  E}\ }\textbf {\bibinfo {volume} {58}},\ \bibinfo {pages} {4828} (\bibinfo
  {year} {1998})}\BibitemShut {NoStop}%
\bibitem [{\citenamefont {Cavagna}\ \emph
  {et~al.}(2019{\natexlab{a}})\citenamefont {Cavagna}, \citenamefont {Culla},
  \citenamefont {Di~Carlo}, \citenamefont {Giardina},\ and\ \citenamefont
  {Grigera}}]{cavagna2019low}%
  \BibitemOpen
  \bibfield  {author} {\bibinfo {author} {\bibfnamefont {A.}~\bibnamefont
  {Cavagna}}, \bibinfo {author} {\bibfnamefont {A.}~\bibnamefont {Culla}},
  \bibinfo {author} {\bibfnamefont {L.}~\bibnamefont {Di~Carlo}}, \bibinfo
  {author} {\bibfnamefont {I.}~\bibnamefont {Giardina}}, \ and\ \bibinfo
  {author} {\bibfnamefont {T.~S.}\ \bibnamefont {Grigera}},\ }\href@noop {}
  {\bibfield  {journal} {\bibinfo  {journal} {Comptes Rendus. Physique}\
  }\textbf {\bibinfo {volume} {20}},\ \bibinfo {pages} {319} (\bibinfo {year}
  {2019}{\natexlab{a}})}\BibitemShut {NoStop}%
\bibitem [{\citenamefont {Cavagna}\ \emph
  {et~al.}(2022{\natexlab{a}})\citenamefont {Cavagna}, \citenamefont {Culla},
  \citenamefont {Feng}, \citenamefont {Giardina}, \citenamefont {Grigera},
  \citenamefont {Kion-Crosby}, \citenamefont {Melillo}, \citenamefont
  {Pisegna}, \citenamefont {Postiglione},\ and\ \citenamefont
  {Villegas}}]{cavagna2022marginal}%
  \BibitemOpen
  \bibfield  {author} {\bibinfo {author} {\bibfnamefont {A.}~\bibnamefont
  {Cavagna}}, \bibinfo {author} {\bibfnamefont {A.}~\bibnamefont {Culla}},
  \bibinfo {author} {\bibfnamefont {X.}~\bibnamefont {Feng}}, \bibinfo {author}
  {\bibfnamefont {I.}~\bibnamefont {Giardina}}, \bibinfo {author}
  {\bibfnamefont {T.~S.}\ \bibnamefont {Grigera}}, \bibinfo {author}
  {\bibfnamefont {W.}~\bibnamefont {Kion-Crosby}}, \bibinfo {author}
  {\bibfnamefont {S.}~\bibnamefont {Melillo}}, \bibinfo {author} {\bibfnamefont
  {G.}~\bibnamefont {Pisegna}}, \bibinfo {author} {\bibfnamefont
  {L.}~\bibnamefont {Postiglione}}, \ and\ \bibinfo {author} {\bibfnamefont
  {P.}~\bibnamefont {Villegas}},\ }\href@noop {} {\bibfield  {journal}
  {\bibinfo  {journal} {Nature Communications}\ }\textbf {\bibinfo {volume}
  {13}},\ \bibinfo {pages} {2315} (\bibinfo {year}
  {2022}{\natexlab{a}})}\BibitemShut {NoStop}%
\bibitem [{\citenamefont {Cavagna}\ \emph
  {et~al.}(2022{\natexlab{b}})\citenamefont {Cavagna}, \citenamefont {Culla},\
  and\ \citenamefont {Grigera}}]{cavagna2022renormalization}%
  \BibitemOpen
  \bibfield  {author} {\bibinfo {author} {\bibfnamefont {A.}~\bibnamefont
  {Cavagna}}, \bibinfo {author} {\bibfnamefont {A.}~\bibnamefont {Culla}}, \
  and\ \bibinfo {author} {\bibfnamefont {T.~S.}\ \bibnamefont {Grigera}},\
  }\href@noop {} {\bibfield  {journal} {\bibinfo  {journal} {Physical Review
  E}\ }\textbf {\bibinfo {volume} {106}},\ \bibinfo {pages} {054136} (\bibinfo
  {year} {2022}{\natexlab{b}})}\BibitemShut {NoStop}%
\bibitem [{\citenamefont {Fermi}\ \emph {et~al.}(1955)\citenamefont {Fermi},
  \citenamefont {Pasta}, \citenamefont {Ulam},\ and\ \citenamefont
  {Tsingou}}]{fermi1955studies}%
  \BibitemOpen
  \bibfield  {author} {\bibinfo {author} {\bibfnamefont {E.}~\bibnamefont
  {Fermi}}, \bibinfo {author} {\bibfnamefont {P.}~\bibnamefont {Pasta}},
  \bibinfo {author} {\bibfnamefont {S.}~\bibnamefont {Ulam}}, \ and\ \bibinfo
  {author} {\bibfnamefont {M.}~\bibnamefont {Tsingou}},\ }\href@noop {} {\emph
  {\bibinfo {title} {Studies of the nonlinear problems}}},\ \bibinfo {type}
  {Tech. Rep.}\ (\bibinfo  {institution} {Los Alamos National Laboratory
  (LANL), Los Alamos, NM (United States)},\ \bibinfo {year} {1955})\BibitemShut
  {NoStop}%
\bibitem [{\citenamefont {Gallavotti}(2008)}]{gallavotti2008fermi}%
  \BibitemOpen
  \bibfield  {author} {\bibinfo {author} {\bibfnamefont {G.}~\bibnamefont
  {Gallavotti}},\ }\href@noop {} {\emph {\bibinfo {title} {The Fermi-Pasta-Ulam
  problem: a status report}}},\ Vol.\ \bibinfo {volume} {728}\ (\bibinfo
  {publisher} {Springer},\ \bibinfo {year} {2008})\BibitemShut {NoStop}%
\bibitem [{\citenamefont {Friesecke}\ and\ \citenamefont
  {Wattis}(1994)}]{friesecke1994existence}%
  \BibitemOpen
  \bibfield  {author} {\bibinfo {author} {\bibfnamefont {G.}~\bibnamefont
  {Friesecke}}\ and\ \bibinfo {author} {\bibfnamefont {J.~A.}\ \bibnamefont
  {Wattis}},\ }\href@noop {} {\bibfield  {journal} {\bibinfo  {journal}
  {Communications in mathematical physics}\ }\textbf {\bibinfo {volume}
  {161}},\ \bibinfo {pages} {391} (\bibinfo {year} {1994})}\BibitemShut
  {NoStop}%
\bibitem [{\citenamefont {Friesecke}\ and\ \citenamefont
  {Matthies}(2003)}]{Friesecke2003geometric}%
  \BibitemOpen
  \bibfield  {author} {\bibinfo {author} {\bibfnamefont {G.}~\bibnamefont
  {Friesecke}}\ and\ \bibinfo {author} {\bibfnamefont {K.}~\bibnamefont
  {Matthies}},\ }\href {\doibase 10.3934/dcdsb.2003.3.105} {\bibfield
  {journal} {\bibinfo  {journal} {Discrete and Continuous Dynamical Systems -
  B}\ }\textbf {\bibinfo {volume} {3}},\ \bibinfo {pages} {105} (\bibinfo
  {year} {2003})}\BibitemShut {NoStop}%
\bibitem [{\citenamefont {Chen}\ and\ \citenamefont
  {Herrmann}(2018)}]{Chen2018Discrete}%
  \BibitemOpen
  \bibfield  {author} {\bibinfo {author} {\bibfnamefont {F.}~\bibnamefont
  {Chen}}\ and\ \bibinfo {author} {\bibfnamefont {M.}~\bibnamefont
  {Herrmann}},\ }\href {\doibase 10.3934/dcds.2018095} {\bibfield  {journal}
  {\bibinfo  {journal} {Discrete and Continuous Dynamical Systems}\ }\textbf
  {\bibinfo {volume} {38}},\ \bibinfo {pages} {2305} (\bibinfo {year}
  {2018})}\BibitemShut {NoStop}%
\bibitem [{\citenamefont {Orlowski}\ and\ \citenamefont
  {Sobczyk}(1989)}]{ORLOWSKI1989solitons}%
  \BibitemOpen
  \bibfield  {author} {\bibinfo {author} {\bibfnamefont {A.}~\bibnamefont
  {Orlowski}}\ and\ \bibinfo {author} {\bibfnamefont {K.}~\bibnamefont
  {Sobczyk}},\ }\href {\doibase https://doi.org/10.1016/0034-4877(89)90036-0}
  {\bibfield  {journal} {\bibinfo  {journal} {Reports on Mathematical Physics}\
  }\textbf {\bibinfo {volume} {27}},\ \bibinfo {pages} {59} (\bibinfo {year}
  {1989})}\BibitemShut {NoStop}%
\bibitem [{\citenamefont {Purwins}\ \emph {et~al.}(2010)\citenamefont
  {Purwins}, \citenamefont {Bödeker},\ and\ \citenamefont
  {and}}]{Purwins2010dissipative}%
  \BibitemOpen
  \bibfield  {author} {\bibinfo {author} {\bibfnamefont {H.-G.}\ \bibnamefont
  {Purwins}}, \bibinfo {author} {\bibfnamefont {H.}~\bibnamefont {Bödeker}}, \
  and\ \bibinfo {author} {\bibfnamefont {S.~A.}\ \bibnamefont {and}},\
  }\href@noop {} {\bibfield  {journal} {\bibinfo  {journal} {Advances in
  Physics}\ }\textbf {\bibinfo {volume} {59}},\ \bibinfo {pages} {485}
  (\bibinfo {year} {2010})}\BibitemShut {NoStop}%
\bibitem [{\citenamefont {Cavagna}\ \emph
  {et~al.}(2021{\natexlab{a}})\citenamefont {Cavagna}, \citenamefont {Feng},
  \citenamefont {Melillo}, \citenamefont {Parisi}, \citenamefont
  {Postiglione},\ and\ \citenamefont {Villegas}}]{Como}%
  \BibitemOpen
  \bibfield  {author} {\bibinfo {author} {\bibfnamefont {A.}~\bibnamefont
  {Cavagna}}, \bibinfo {author} {\bibfnamefont {X.}~\bibnamefont {Feng}},
  \bibinfo {author} {\bibfnamefont {S.}~\bibnamefont {Melillo}}, \bibinfo
  {author} {\bibfnamefont {L.}~\bibnamefont {Parisi}}, \bibinfo {author}
  {\bibfnamefont {L.}~\bibnamefont {Postiglione}}, \ and\ \bibinfo {author}
  {\bibfnamefont {P.}~\bibnamefont {Villegas}},\ }\href@noop {} {\bibfield
  {journal} {\bibinfo  {journal} {IEEE Transactions on Instrumentation and
  Measurement}\ }\textbf {\bibinfo {volume} {70}},\ \bibinfo {pages} {1}
  (\bibinfo {year} {2021}{\natexlab{a}})}\BibitemShut {NoStop}%
\bibitem [{\citenamefont {Attanasi}\ \emph {et~al.}(2015)\citenamefont
  {Attanasi}, \citenamefont {Cavagna}, \citenamefont {Del~Castello},
  \citenamefont {Giardina}, \citenamefont {Jeli{\'c}}, \citenamefont {Melillo},
  \citenamefont {Parisi}, \citenamefont {Pellacini}, \citenamefont {Shen},
  \citenamefont {Silvestri} \emph {et~al.}}]{attanasi2015greta}%
  \BibitemOpen
  \bibfield  {author} {\bibinfo {author} {\bibfnamefont {A.}~\bibnamefont
  {Attanasi}}, \bibinfo {author} {\bibfnamefont {A.}~\bibnamefont {Cavagna}},
  \bibinfo {author} {\bibfnamefont {L.}~\bibnamefont {Del~Castello}}, \bibinfo
  {author} {\bibfnamefont {I.}~\bibnamefont {Giardina}}, \bibinfo {author}
  {\bibfnamefont {A.}~\bibnamefont {Jeli{\'c}}}, \bibinfo {author}
  {\bibfnamefont {S.}~\bibnamefont {Melillo}}, \bibinfo {author} {\bibfnamefont
  {L.}~\bibnamefont {Parisi}}, \bibinfo {author} {\bibfnamefont
  {F.}~\bibnamefont {Pellacini}}, \bibinfo {author} {\bibfnamefont
  {E.}~\bibnamefont {Shen}}, \bibinfo {author} {\bibfnamefont {E.}~\bibnamefont
  {Silvestri}},  \emph {et~al.},\ }\href@noop {} {\bibfield  {journal}
  {\bibinfo  {journal} {IEEE transactions on pattern analysis and machine
  intelligence}\ }\textbf {\bibinfo {volume} {37}},\ \bibinfo {pages} {2451}
  (\bibinfo {year} {2015})}\BibitemShut {NoStop}%
\bibitem [{\citenamefont {Cavagna}\ \emph {et~al.}(2018)\citenamefont
  {Cavagna}, \citenamefont {Giardina},\ and\ \citenamefont
  {Grigera}}]{cavagna2018physics}%
  \BibitemOpen
  \bibfield  {author} {\bibinfo {author} {\bibfnamefont {A.}~\bibnamefont
  {Cavagna}}, \bibinfo {author} {\bibfnamefont {I.}~\bibnamefont {Giardina}}, \
  and\ \bibinfo {author} {\bibfnamefont {T.~S.}\ \bibnamefont {Grigera}},\
  }\href@noop {} {\bibfield  {journal} {\bibinfo  {journal} {Physics Reports}\
  }\textbf {\bibinfo {volume} {728}},\ \bibinfo {pages} {1} (\bibinfo {year}
  {2018})}\BibitemShut {NoStop}%
\bibitem [{\citenamefont {Cavagna}\ \emph {et~al.}(2024)\citenamefont
  {Cavagna}, \citenamefont {Crist{\'\i}n}, \citenamefont {Giardina},
  \citenamefont {Grigera},\ and\ \citenamefont {Veca}}]{cavagna2024discrete}%
  \BibitemOpen
  \bibfield  {author} {\bibinfo {author} {\bibfnamefont {A.}~\bibnamefont
  {Cavagna}}, \bibinfo {author} {\bibfnamefont {J.}~\bibnamefont
  {Crist{\'\i}n}}, \bibinfo {author} {\bibfnamefont {I.}~\bibnamefont
  {Giardina}}, \bibinfo {author} {\bibfnamefont {T.~S.}\ \bibnamefont
  {Grigera}}, \ and\ \bibinfo {author} {\bibfnamefont {M.}~\bibnamefont
  {Veca}},\ }\href@noop {} {\bibfield  {journal} {\bibinfo  {journal} {Journal
  of Physics A: Mathematical and Theoretical}\ }\textbf {\bibinfo {volume}
  {57}},\ \bibinfo {pages} {415002} (\bibinfo {year} {2024})}\BibitemShut
  {NoStop}%
\bibitem [{\citenamefont {Allen}\ and\ \citenamefont
  {Tildesley}(1987)}]{Allen1987}%
  \BibitemOpen
  \bibfield  {author} {\bibinfo {author} {\bibfnamefont {M.~P.}\ \bibnamefont
  {Allen}}\ and\ \bibinfo {author} {\bibfnamefont {D.~J.}\ \bibnamefont
  {Tildesley}},\ }\href@noop {} {\emph {\bibinfo {title} {Computer
  {{Simulation}} of {{Liquids}}}}}\ (\bibinfo  {publisher} {{Clarendon
  Press}},\ \bibinfo {address} {Oxford},\ \bibinfo {year} {1987})\BibitemShut
  {NoStop}%
\bibitem [{\citenamefont {Swope}\ \emph {et~al.}(1982)\citenamefont {Swope},
  \citenamefont {Andersen}, \citenamefont {Berens},\ and\ \citenamefont
  {Wilson}}]{swope_computer_1982-1}%
  \BibitemOpen
  \bibfield  {author} {\bibinfo {author} {\bibfnamefont {W.~C.}\ \bibnamefont
  {Swope}}, \bibinfo {author} {\bibfnamefont {H.~C.}\ \bibnamefont {Andersen}},
  \bibinfo {author} {\bibfnamefont {P.~H.}\ \bibnamefont {Berens}}, \ and\
  \bibinfo {author} {\bibfnamefont {K.~R.}\ \bibnamefont {Wilson}},\ }\href
  {\doibase 10.1063/1.442716} {\bibfield  {journal} {\bibinfo  {journal} {The
  Journal of Chemical Physics}\ }\textbf {\bibinfo {volume} {76}},\ \bibinfo
  {pages} {637} (\bibinfo {year} {1982})}\BibitemShut {NoStop}%
\bibitem [{\citenamefont {Cavagna}\ \emph
  {et~al.}(2021{\natexlab{b}})\citenamefont {Cavagna}, \citenamefont
  {Di~Carlo}, \citenamefont {Giardina}, \citenamefont {Grigera},\ and\
  \citenamefont {Pisegna}}]{cavagna2021equilibrium}%
  \BibitemOpen
  \bibfield  {author} {\bibinfo {author} {\bibfnamefont {A.}~\bibnamefont
  {Cavagna}}, \bibinfo {author} {\bibfnamefont {L.}~\bibnamefont {Di~Carlo}},
  \bibinfo {author} {\bibfnamefont {I.}~\bibnamefont {Giardina}}, \bibinfo
  {author} {\bibfnamefont {T.~S.}\ \bibnamefont {Grigera}}, \ and\ \bibinfo
  {author} {\bibfnamefont {G.}~\bibnamefont {Pisegna}},\ }\href@noop {}
  {\bibfield  {journal} {\bibinfo  {journal} {Physical Review Research}\
  }\textbf {\bibinfo {volume} {3}},\ \bibinfo {pages} {013210} (\bibinfo {year}
  {2021}{\natexlab{b}})}\BibitemShut {NoStop}%
\bibitem [{\citenamefont {Cavagna}\ \emph {et~al.}(2023)\citenamefont
  {Cavagna}, \citenamefont {Di~Carlo}, \citenamefont {Giardina}, \citenamefont
  {Grigera}, \citenamefont {Melillo}, \citenamefont {Parisi}, \citenamefont
  {Pisegna},\ and\ \citenamefont {Scandolo}}]{cavagna2023natural}%
  \BibitemOpen
  \bibfield  {author} {\bibinfo {author} {\bibfnamefont {A.}~\bibnamefont
  {Cavagna}}, \bibinfo {author} {\bibfnamefont {L.}~\bibnamefont {Di~Carlo}},
  \bibinfo {author} {\bibfnamefont {I.}~\bibnamefont {Giardina}}, \bibinfo
  {author} {\bibfnamefont {T.~S.}\ \bibnamefont {Grigera}}, \bibinfo {author}
  {\bibfnamefont {S.}~\bibnamefont {Melillo}}, \bibinfo {author} {\bibfnamefont
  {L.}~\bibnamefont {Parisi}}, \bibinfo {author} {\bibfnamefont
  {G.}~\bibnamefont {Pisegna}}, \ and\ \bibinfo {author} {\bibfnamefont
  {M.}~\bibnamefont {Scandolo}},\ }\href@noop {} {\bibfield  {journal}
  {\bibinfo  {journal} {Nature Physics}\ }\textbf {\bibinfo {volume} {19}},\
  \bibinfo {pages} {1043} (\bibinfo {year} {2023})}\BibitemShut {NoStop}%
\bibitem [{\citenamefont {Zinn-Justin}(2002)}]{zinnjustin_QFTCF}%
  \BibitemOpen
  \bibfield  {author} {\bibinfo {author} {\bibfnamefont {J.}~\bibnamefont
  {Zinn-Justin}},\ }\href {https://hal.archives-ouvertes.fr/hal-00120423}
  {\enquote {\bibinfo {title} {{Quantum Field Theory and Critical
  Phenomena}},}\ } (\bibinfo {year} {2002}),\ \bibinfo {note} {international
  Series of Monographs on Physics 113, 1054 pp. (2002), Fourth
  Edition.}\BibitemShut {Stop}%
\bibitem [{\citenamefont {Cavagna}\ \emph
  {et~al.}(2015{\natexlab{b}})\citenamefont {Cavagna}, \citenamefont
  {Giardina}, \citenamefont {Grigera}, \citenamefont {Jelic}, \citenamefont
  {Levine}, \citenamefont {Ramaswamy},\ and\ \citenamefont
  {Viale}}]{cavagna2015silent}%
  \BibitemOpen
  \bibfield  {author} {\bibinfo {author} {\bibfnamefont {A.}~\bibnamefont
  {Cavagna}}, \bibinfo {author} {\bibfnamefont {I.}~\bibnamefont {Giardina}},
  \bibinfo {author} {\bibfnamefont {T.~S.}\ \bibnamefont {Grigera}}, \bibinfo
  {author} {\bibfnamefont {A.}~\bibnamefont {Jelic}}, \bibinfo {author}
  {\bibfnamefont {D.}~\bibnamefont {Levine}}, \bibinfo {author} {\bibfnamefont
  {S.}~\bibnamefont {Ramaswamy}}, \ and\ \bibinfo {author} {\bibfnamefont
  {M.}~\bibnamefont {Viale}},\ }\href@noop {} {\bibfield  {journal} {\bibinfo
  {journal} {Physical Review Letters}\ }\textbf {\bibinfo {volume} {114}},\
  \bibinfo {pages} {218101} (\bibinfo {year} {2015}{\natexlab{b}})}\BibitemShut
  {NoStop}%
\bibitem [{\citenamefont {Cavagna}\ \emph
  {et~al.}(2019{\natexlab{b}})\citenamefont {Cavagna}, \citenamefont
  {Di~Carlo}, \citenamefont {Giardina}, \citenamefont {Grandinetti},
  \citenamefont {Grigera},\ and\ \citenamefont
  {Pisegna}}]{cavagna2019renormalizationl}%
  \BibitemOpen
  \bibfield  {author} {\bibinfo {author} {\bibfnamefont {A.}~\bibnamefont
  {Cavagna}}, \bibinfo {author} {\bibfnamefont {L.}~\bibnamefont {Di~Carlo}},
  \bibinfo {author} {\bibfnamefont {I.}~\bibnamefont {Giardina}}, \bibinfo
  {author} {\bibfnamefont {L.}~\bibnamefont {Grandinetti}}, \bibinfo {author}
  {\bibfnamefont {T.~S.}\ \bibnamefont {Grigera}}, \ and\ \bibinfo {author}
  {\bibfnamefont {G.}~\bibnamefont {Pisegna}},\ }\href@noop {} {\bibfield
  {journal} {\bibinfo  {journal} {Physical Review E}\ }\textbf {\bibinfo
  {volume} {100}},\ \bibinfo {pages} {062130} (\bibinfo {year}
  {2019}{\natexlab{b}})}\BibitemShut {NoStop}%
\bibitem [{\citenamefont {Cavagna}\ \emph
  {et~al.}(2019{\natexlab{c}})\citenamefont {Cavagna}, \citenamefont
  {Di~Carlo}, \citenamefont {Giardina}, \citenamefont {Grandinetti},
  \citenamefont {Grigera},\ and\ \citenamefont
  {Pisegna}}]{cavagna2019dynamical}%
  \BibitemOpen
  \bibfield  {author} {\bibinfo {author} {\bibfnamefont {A.}~\bibnamefont
  {Cavagna}}, \bibinfo {author} {\bibfnamefont {L.}~\bibnamefont {Di~Carlo}},
  \bibinfo {author} {\bibfnamefont {I.}~\bibnamefont {Giardina}}, \bibinfo
  {author} {\bibfnamefont {L.}~\bibnamefont {Grandinetti}}, \bibinfo {author}
  {\bibfnamefont {T.~S.}\ \bibnamefont {Grigera}}, \ and\ \bibinfo {author}
  {\bibfnamefont {G.}~\bibnamefont {Pisegna}},\ }\href@noop {} {\bibfield
  {journal} {\bibinfo  {journal} {Physical Review Letters}\ }\textbf {\bibinfo
  {volume} {123}},\ \bibinfo {pages} {268001} (\bibinfo {year}
  {2019}{\natexlab{c}})}\BibitemShut {NoStop}%
\bibitem [{\citenamefont {Tu}\ \emph {et~al.}(1998{\natexlab{b}})\citenamefont
  {Tu}, \citenamefont {Toner},\ and\ \citenamefont {Ulm}}]{toner+al_98}%
  \BibitemOpen
  \bibfield  {author} {\bibinfo {author} {\bibfnamefont {Y.}~\bibnamefont
  {Tu}}, \bibinfo {author} {\bibfnamefont {J.}~\bibnamefont {Toner}}, \ and\
  \bibinfo {author} {\bibfnamefont {M.}~\bibnamefont {Ulm}},\ }\href {\doibase
  10.1103/PhysRevLett.80.4819} {\bibfield  {journal} {\bibinfo  {journal}
  {Phys. Rev. Lett.}\ }\textbf {\bibinfo {volume} {80}},\ \bibinfo {pages}
  {4819} (\bibinfo {year} {1998}{\natexlab{b}})}\BibitemShut {NoStop}%
\bibitem [{\citenamefont {Ryder}(1996)}]{ryder1996quantum}%
  \BibitemOpen
  \bibfield  {author} {\bibinfo {author} {\bibfnamefont {L.~H.}\ \bibnamefont
  {Ryder}},\ }\href@noop {} {\emph {\bibinfo {title} {Quantum field theory}}}\
  (\bibinfo  {publisher} {Cambridge university press},\ \bibinfo {year}
  {1996})\BibitemShut {NoStop}%
\bibitem [{\citenamefont {Patashinskii}\ and\ \citenamefont
  {Pokrovskii}(1973)}]{pata1973chi_longi}%
  \BibitemOpen
  \bibfield  {author} {\bibinfo {author} {\bibfnamefont {A.~Z.}\ \bibnamefont
  {Patashinskii}}\ and\ \bibinfo {author} {\bibfnamefont {V.~L.}\ \bibnamefont
  {Pokrovskii}},\ }\href@noop {} {\bibfield  {journal} {\bibinfo  {journal}
  {Zh. Eksp. Teor. Fiz.}\ }\textbf {\bibinfo {volume} {64}},\ \bibinfo {pages}
  {1445} (\bibinfo {year} {1973})}\BibitemShut {NoStop}%
\bibitem [{\citenamefont {Patashinskii}\ and\ \citenamefont
  {Pokrovskii}(1979)}]{patashinskii_book}%
  \BibitemOpen
  \bibfield  {author} {\bibinfo {author} {\bibfnamefont {A.~Z.}\ \bibnamefont
  {Patashinskii}}\ and\ \bibinfo {author} {\bibfnamefont {V.~L.}\ \bibnamefont
  {Pokrovskii}},\ }\href@noop {} {\emph {\bibinfo {title} {Fluctuation Theory
  of Phase Transitions}}}\ (\bibinfo  {publisher} {Pergamon Press},\ \bibinfo
  {year} {1979})\BibitemShut {NoStop}%
\end{thebibliography}%

%%%%%%%%%%%%%%%%%%%%%%%%%%%%%%%%%%%%%%%%%%%%%%%%%%%%%%%%%%%%%%%%%
%%%%%%%%%%%%%%%%%%%%%%%%%%%%%%%%%%%%%%%%%%%%%%%%%%%%%%%%%%%%%%%%%
%%%%%%%%%%%%%%%%%%%%%%%%%%%%%%%%%%%%%%%%%%%%%%%%%%%%%%%%%%%%%%%%%

\section{Video Captions}

\noindent
{\bf Video S1: A real flock turning.} 
A flock of starlings turning over their roosting site, at Piazza dei Cinquecento, in Rome. Images are taken at 10 frames-per-second by one of the three cameras of the stereoscopic system.

\vskip 0.5 truecm

\noindent
{\bf Video S2: A real flock turning -- 3D reconstruction.} 
After tracking and stereo matching, the 3D positions of all birds belonging to the flock in Video S1 are reconstructed. Wing-flapping is evident from the zig-zag trajectories of the birds. Particles are turned from grey to red once their heading changes beyond some threshold, highlighting  the underdamped propagation of the phase wave. Note that -- due to the upward-tilted positions of the three cameras -- no axis in this reference frame coincides with gravity.

\vskip 0.5 truecm

\noindent
{\bf Video S3: ISM Simulations -- Underdamped regime.} 
Numerical simulations of the $3d$ ISM in open boundary conditions (Fig.1a). The initiator particle (cyan) is controlled and turned by 90 degrees. Thanks to inertia, in the underdamped regime there is linear propagation of the phase disturbances, therefore the initiator's perturbation elicit a collective turn of the group.

\vskip 0.5 truecm

\noindent
{\bf Video S4: ISM Simulations -- Overdamped regime.} 
Numerical simulations of the $3d$ ISM in open boundary conditions (Fig.1c). The initiator particle (cyan) is controlled and turned by 90 degrees. In the overdamped regime inertia becomes irrelevant and friction dominates; as a consequence, the perturbation produced by the initiator does not elicit a collective turn, and the group loses cohesion.

\vskip 0.5 truecm

\noindent
{\bf Video S5: ISM+FPUT Simulations.} 
Numerical simulations of the $3d$ ISM+FPUT in open boundary conditions (Fig.3a). The initiator particle (cyan) is controlled and turned by 90 degrees. Despite the fact that spontaneous phase fluctuations are overdamped in this model, the finite phase disturbance produced by the initiator's change of direction pushes the effective stiffness into the underdamped regime, hence yielding a collective turn.

%%%%%%%%%%%%%%%%%%%%%%%%%%%%%%%%%%%%%%%%%%%%%%%%%%%%%%%%%%%%%%%%%
%%%%%%%%%%%%%%%%%%%%%%%%%%%%%%%%%%%%%%%%%%%%%%%%%%%%%%%%%%%%%%%%%
%%%%%%%%%%%%%%%%%%%%%%%%%%%%%%%%%%%%%%%%%%%%%%%%%%%%%%%%%%%%%%%%%

\section{Supplemental Material}
%% For the Supplemental Material I reset figure counting
\renewcommand{\thefigure}{S\arabic{figure}}
\setcounter{figure}{0}

\subsection{New experiments}

In standard physical systems, under linear response theory, we can normally compute correlation and response during the same perturbation experiment, because the effect of the perturbation on the correlation function is of higher order. In the case of real flocks, though, doing 
that is unwise: nonlinear effects may be (in fact, are) very strong, hence computing the correlation function during a perturbation, that is during a collective turn, we might be mixing several confounding effects. For this reason in this work we want to compare the physics of the unperturbed correlation function, with the propagation of phase waves during collective turns, i.e. during a perturbation. From the experimental point of view, this is not straightforward.

The original experiments on turning flocks conducted in \cite{attanasi+al_14} were performed with fixed cameras; with that setup, it was possible to record for a reasonably long time those few flocks (just 12 in \cite{attanasi+al_14}) that happened to turn right in front of the cameras, thus staying in the field of view for long, despite their high speed. But a flock flying straight (no turn, no perturbation) stays in the field of view of all cameras in the stereoscopic system for just 2-3 seconds; this is insufficient to have an accurate determination of the correlation function, as the frequency step is simply too large with such short an acquisition and any potential spin-wave peak may be lost between the large frequency bins. Therefore, to calculate the correlation function we had to develop a new experimental setup, in which cameras are able to pan and follow the moving flock.

Three high--speed synchronized cameras (IDT OS10-4K; resolution $3840 px \times 2400 px$; sensor size $17.9 mm \times 11.2 mm$), equipped with Schneider Xenoplan 28 mm f/2.0 lenses, are coupled with high--speed one--axis rotational stages (Newport RVS80CC; nominal accuracy $10^{-4} rad$; nominal home repeatability $4\cdot 10^{-3} rad$) that allow each camera to rotate on the plane orthogonal to its sensor. The direction and speed of rotation are manually controlled by an operator via a joypad (Logitech F310) connected to the three stages, which rotate synchronously and with the same parameters. This new panning system has been validated and extensively tested in terms of $3D$ reconstruction accuracy; see \cite{Como} for further details on the reconstruction method and hardware/software specifications of the apparatus.
The new experimental setup allowed us to have acquisitions as long as $12.5$ seconds, a very significant advancement compared to the previous apparatus, that produced acquisitions of $3$ seconds max.

Data were collected from Palazzo Massimo alle Terme, Piazza dei Cinquecento (Rome - Italy), in different experimental campaigns between $2019$ and $2023$, with flocks at a typical distance from the cameras between $80m$ and $130m$ and system baseline of $25m$; typical rotational speed is below $4^\circ/s$. Tracking is performed using the software GReTA \cite{attanasi2015greta}, adapted to cope with camera rotation.

\subsection{The spatio-temporal velocity correlation function} 

The connected velocity correlation function is defined as,
\begin{equation}
    C(\boldsymbol{k},t) =   \frac{1}{N}  \left\langle \sum_{i,j}^N \delta\bv_i(t_0)\cdot\delta\bv_j(t_0+t) e^{-i\boldsymbol{k}\cdot \boldsymbol{r}_{ij}}\right\rangle_{t_0} \ ,
\label{eq:ckt_general}
\end{equation}
where the velocity fluctuation of particle $i$ is given by $\delta \bv_i(t) = \bv_i(t)-(1/N)\sum_k\bv_k(t)$. The relative displacement between particles $i$ and $j$ is given by $\boldsymbol{r}_{ij}=\boldsymbol{r}^{CM}_i(t_0) - \boldsymbol{r}^{CM}_j(t_0+t)$, where the superscript indicates positions measured with respect to the instantaneous center of mass. The average $\langle\cdot\rangle_{t_0}$ is taken over the time $t_0$. To properly account for periodic boundary conditions in the numerical simulations, particle trajectories are first unfolded prior to the transformation to the center of mass frame. The unfolded position is computed as $\boldsymbol{r}_i=\boldsymbol{r}_i^{box}+(n_x,n_y,n_z)L$, where $\boldsymbol{r}_{i}^{box}$ denotes the position of particle $i$ within the cubic domain, $n_\alpha$ counts the number of crossings of particle $i$ along direction $\alpha$, and $L$ denotes the cubic box length.

To improve statistical accuracy, particularly relevant for the experimental analysis, we assume isotropy. Under this assumption, the correlation function $C(\boldsymbol{k}, t)$ becomes independent of the orientation of the wavevector and can thus be written as $C(k,t)$, where $k = |\boldsymbol{k}|$. To obtain an explicitly isotropic expression, we average over the direction of $\boldsymbol{k}$. In experimental configurations, which exhibit continuous rotational invariance, this averaging corresponds to a spherical average over all orientations of the wavevector. This yields the following expression for the isotropic correlation function (see 
\cite{cavagna2018physics}),
\begin{equation}
    C(k,t) =   \frac{1}{N}  \left\langle \sum_{i,j}^N \delta\bv_i(t_0)\cdot\delta\bv_j(t_0+t) \frac{\sin \left( k r_{ij} \right)}{k r_{ij}}  \right\rangle_{t_0} \ ,
    \label{eq:ckt_exp} 
\end{equation}
with $r_{ij}=\lvert \boldsymbol{r}_{ij} \rvert$. 
On the other hand, numerical simulations are performed in a cubic domain, which only preserves a discrete rotational symmetry. In this case, we average over the  Cartesian directions:
\begin{equation}
    \label{eq:ckt_sim}
    \begin{split}
        C(k,t) &= \frac{1}{6} \Big( C(\boldsymbol{k}_x,t) + C(\boldsymbol{k}_y,t) + C(\boldsymbol{k}_z,t) \\
               &\quad + C(-\boldsymbol{k}_x,t) + C(-\boldsymbol{k}_y,t) + C(-\boldsymbol{k}_z,t) \Big),
    \end{split}
\end{equation}
where $\boldsymbol{k}_x=(k,0,0)$, $\boldsymbol{k}_y=(0,k,0)$ and $\boldsymbol{k}_z=(0,0,k)$. 
Finally, we calculate the correlation function $C(k,\omega)$  as the temporal Fourier transform of $C(k,t)$.

\subsubsection{Wing-flapping peaks are not spin-wave peaks}

Starlings flap their wings with period $T_\mathrm{flap} \approx 0.1$ sec. At variance with \cite{attanasi+al_14}, where we worked in the domain of time, here we work in the domain of frequency, and we do not filter out this oscillation, as this could be risky, given that we are looking for the spin-wave peaks of the correlation function. The wing-flapping peaks can be clearly seen in the experimental correlation function, right at $\omega = 2\pi/T_\mathrm{flap} \approx 68$ Hz (Fig.S1). To clear the temptation to interpret these as the spin-wave peaks we were looking for,  it is sufficient to note that their position {\it does not} change with $k$. Moreover, spin-wave peaks -- if existent -- should be found at a different frequency: direct estimated of the speed of propagation of traveling waves in starling flocks, gives $c_s \in [10:20]$ m/s \cite{attanasi+al_14}, which at $k=3$ m$^{-1}$ would produce a SW peak at $\omega \in [30:60]$ Hz, which is not compatible with this correlation function. Finally, the clear central peak -- completely absent in the inertial underdamped theory -- is an unmistakable sign of Lorentzian, instead of spin-wave, correlation.

%%%%%%%%%%%%%%%%%%%%%%%%%%%%%%%%%%%%%%%%%%
\begin{figure}
\centering
\includegraphics[width=0.4 \textwidth]{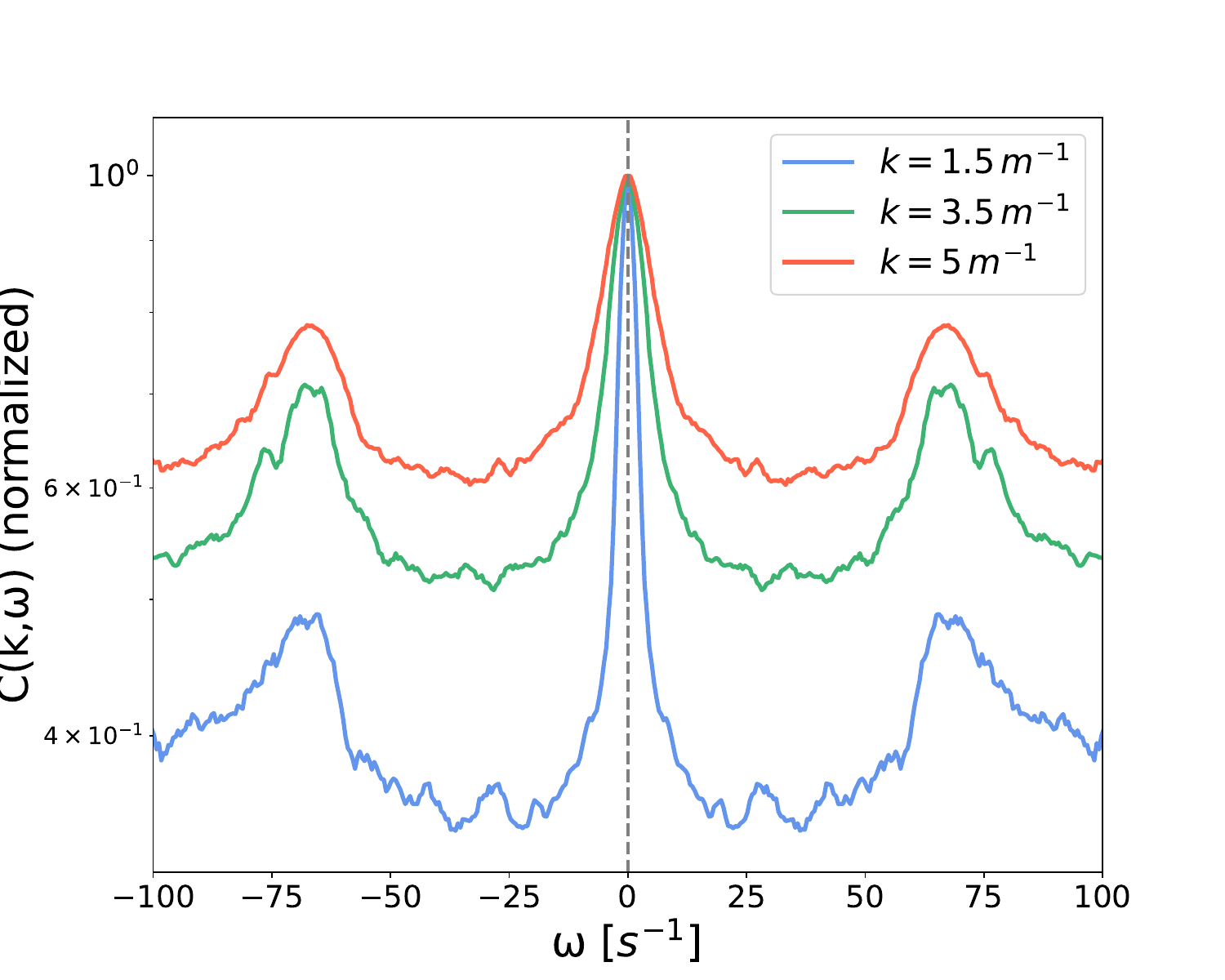}
\caption{{\bf Wing-flapping peaks.} These peaks (enhanced here by the logarithmic scale in $y$) show up at $\omega\approx 68$ Hz, consistent with the physiological wing-flapping frequency of starlings, $\nu_\mathrm{flap}  \approx 10$ Hz. The peaks' position does not change by changing $k$, a clear proof that these are {\it not} spin-wave peaks.}
\end{figure}
%%%%%%%%%%%%%%%%%%%%%%%%%%%%%%%%%%%%%%%%%%%
\subsubsection{Experimental correlation functions}

\begin{figure*}[ht]
	\centering
	\includegraphics[width=0.87\textwidth]{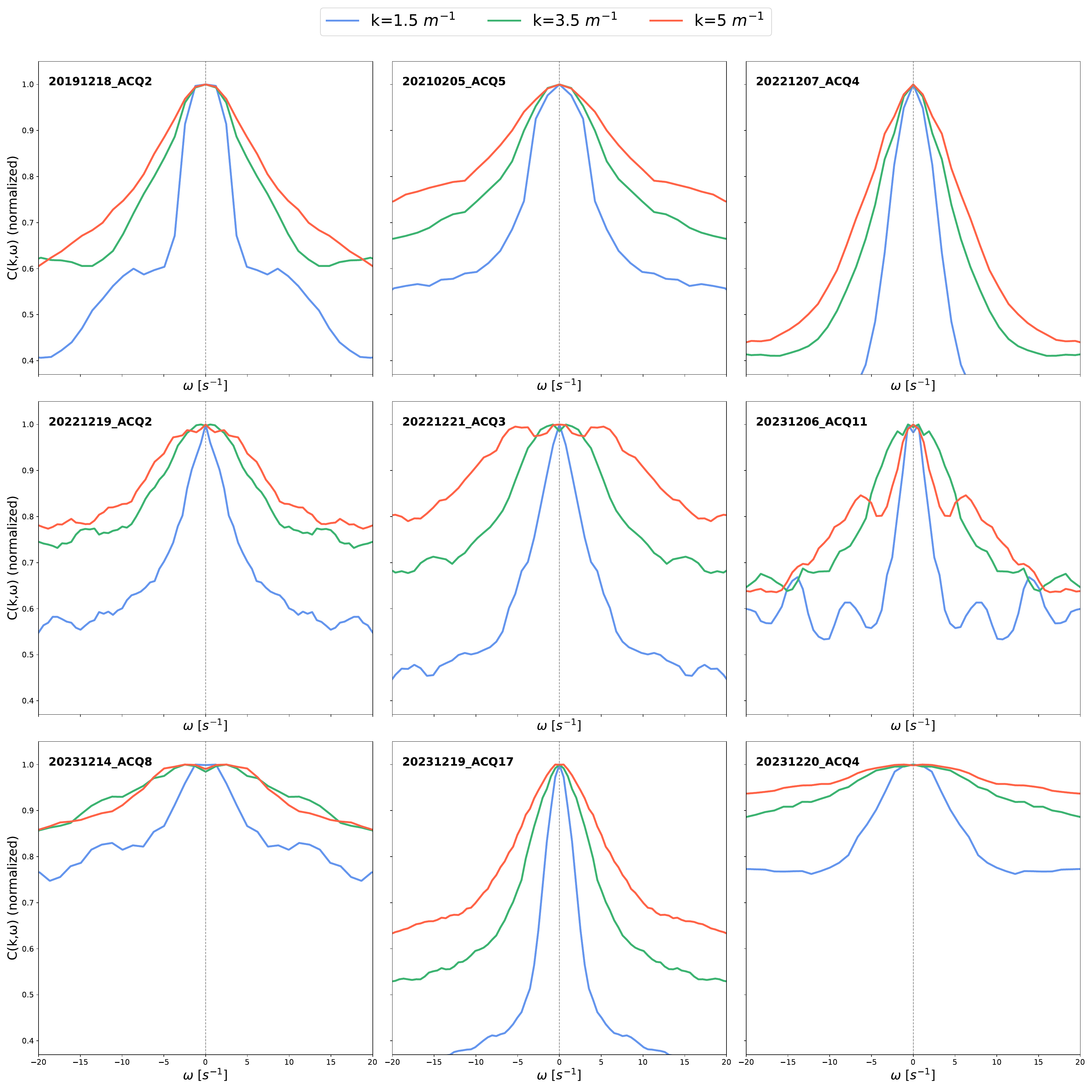}
	\caption{Correlation functions $C(k,\omega)$ of the nine flocks analyzed. In all the experimental flocks, the correlation function has a quasi-Lorentzian form, with no spin-wave peaks, and an half-width frequency scaling as $\omega_L \sim k^2$, as in the overdamped phase of the ISM (Fig.1d).}
	\label{FIG6}
\end{figure*}

In this section we present the correlation functions of several flocks, in addition to the one shown in the main text (Fig. 2f). As is clear from Fig. \ref{FIG6}, in all cases no spin-wave peaks are observed and each correlation function exhibits a purely Lorentzian shape. The parameters of the analyzed flocks are listed in Table \ref{table1}.

\begin{table}[ht]
	\centering
	\begin{ruledtabular}
		\begin{tabular}{lll}
			Flock & $N$ & $T_{MAX}[s]$ \\ 
			\hline
			20191218\_ACQ2 & 875 & 5.10 \\
			20210205\_ACQ5 & 1514 & 4.45 \\
			20221207\_ACQ4 & 180 & 5.52 \\
			20221219\_ACQ2 & 220 & 11.33 \\
			20221221\_ACQ3 & 312 & 8.33 \\
			20231206\_ACQ11 & 249 & 9.99 \\
			20231214\_ACQ8 & 598 & 5.06 \\
			20231219\_ACQ17 & 958 & 12.47 \\
			20231220\_ACQ4 & 431 & 5.67 \\
		\end{tabular}
		
	\end{ruledtabular}
	
	\caption{Table with the number of trajectories $N$ and the acquisition time $T_{MAX}$ for each flock. In Fig. 2f of the main text Flock 20231219\_ACQ17 is shown.}
	\label{table1}
\end{table}

\subsubsection{Characteristic frequency vs wave vector in the various cases}

\begin{figure}
\centering
\includegraphics[width=0.45 \textwidth]{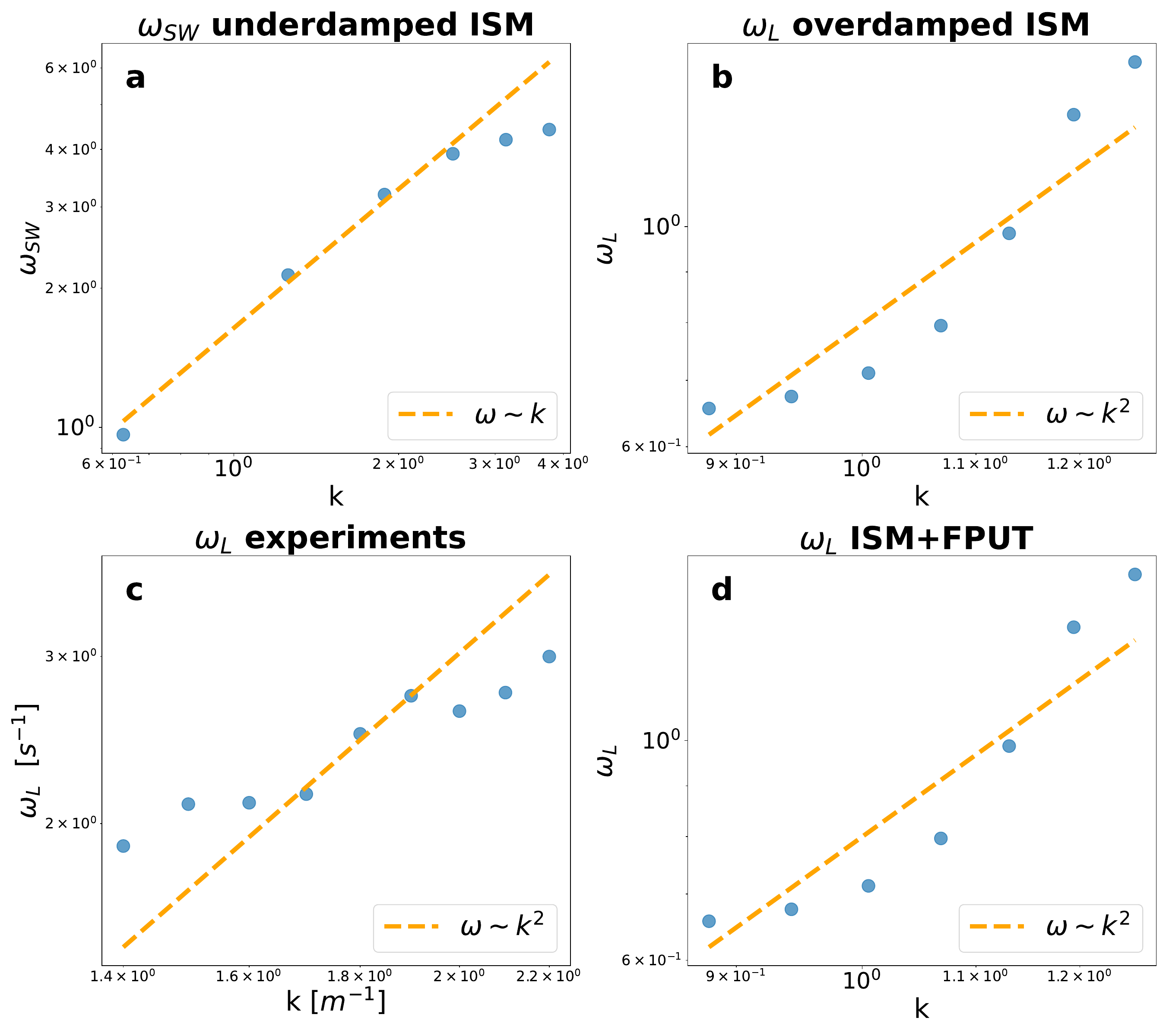}
\caption{{\bf Characteristic frequency.} The characteristic frequency of the various cases is plotted against the wave-vector $k$ on a log-log scale.}
\label{FIG5}
\end{figure}

In all Lorentzian cases, simulations and experiments, the characteristic frequency, $\omega_\mathrm{L}$, has been calculated as the half-width of the central peak at $\omega=0$ of the correlation function, $C(k,\omega)$. In the case of spin-wave form (ISM-underdamped), the characteristic frequency, $\omega_\mathrm{SW}$, has been calculated as the position of the spin-wave peak. The dependence of the various characteristic frequencies on $k$ are reported in Fig.~\ref{FIG5}. In the spin-wave case of underdamped ISM we expect $\omega_\mathrm{SW}(k)$ to be linear for small $k$, and to deviate for larger $k$, due to the discrete nature of the (active) lattice, which is the case; it is interesting to note that this deviation with respect to the linear behaviour of the continuous prediction seems very similar to the equilibrium case, exactly calculated in \cite{cavagna2024discrete} for the fully conservative case. In the ISM-overdamped case, as well as in real experiment and in the ISM+FPUT case, the half-width frequency is not far from the continuous theory prediction, $\omega_\mathrm{L} \sim k^2$.

%%%%%%%%%%%%%%%%%%%%%%%%%%%%%%%%%%%%%%%%%%
%%%%%%%%%%%%%%%%%%%%%%%%%%%%%%%%%%%%%%%%%%%

\subsection{Low-temperature expansion in the ordered phase}

To derive the low temperature expansion, it is convenient to rewrite the pair of first order equations  for the spin and the velocity Eqs.~(\ref{ism}), as a single second order equation for the velocity vector only. This can be easily done by taking a further time derivative of the first equation and exploiting the second one to eliminate the spin variable. We obtain,\footnote{
In the following, we are assuming that $\bv_i \cdot {\mathbf s}_i=0$. This assumption, which simplifies computations and formulas, is however not strictly necessary. Any component of the spin longitudinal to the velocity indeed evolves following a close equation that leads in the stationary state to a zero average fluctuating term. It does not contribute to any significant qualitative difference as compared to case we consider here (see - e.g. \cite{cavagna2024discrete}), and it is quantitatively sub-leading in the low-temperature region.
}
\beq
\chi \frac{d^2 \bv_i}{dt^2} = v_0^2 \bF_i^\perp  -\chi \left (\frac{1}{v_0}  \frac{d\bv_i}{dt}\right )^2\bv_i -\eta \frac{d\bv_i}{dt} + {\boldsymbol \xi}_i \times \bv_i \ ,
\label{eqv}
\eeq
where, 
\beq
\bF_i = - \frac{\partial H}{\partial \bv_i} = -\frac{J}{v_0^2}\sum_{ij} n_{ij} (\bv_i-\bv_j) = -\frac{J}{v_0^2}\sum_{ij} \Lambda_{ij} \bv_j  \ ,
\eeq
and $\bF_i^\perp = \bF_i-(\bF_i\cdot\bv_i/v_0)\bv_i/v_0$ is the component of the force perpendicular to the velocity vector. The structure of this equation is determined by the  constraint on the speed, $|\bv_i|=v_0$: only the perpendicular component of the force contributes to the dynamics, there is a  `centripetal' term (second term at the r.h.s.), and the noise has effectively only two independent coordinates. Apart from this, Eq.~(\ref{eqv}) represents a standard underdamped relaxation dynamics in presence of alignment interactions. In the overdamped limit $\chi/\eta \to 0$, it reduces to the standard (continuous time) Vicsek model.
\vskip 0.5 cm

\subsubsection{Low temperature expansion for ISM}

Eq.~(\ref{eqv}) can be further simplified at low temperature. In this region, the system is very ordered and the velocity fluctuations transverse to the global velocity $\boldsymbol V$ are very small. Introducing the unit vector, $\bn\equiv \boldsymbol V/|\boldsymbol V|$, it is then convenient to decompose the velocity as
\beq
\bv_i = \bv_i^L + \bpi_i= v_i^L \bn + \bpi_i = \bn\sqrt{v_0^2 - \bpi_i^2} + \bpi_i\ ,
\eeq
where we exploited the fixed speed constraint. Since $\bpi_i^2 \ll 1$,  we have $v_i^L=\sqrt{v_0^2 - \bpi_i^2}\approx v_0 (1-\bpi_i^2/(2v_0^2)$. The longitudinal velocity component is effectively slave to the transverse one. The dynamical equation for $\bpi_i$ can be found by expanding the r.h.s. of Eq.\eqref{eqv}.  The force term becomes, to linear order,
\begin{align}
&v_0^2 \bF_i^\perp = -J \sum_{j} \Lambda_{ij} \left[ \bv_i -(\bv_j\cdot \bv_i) \frac{\bv_i}{v_0^2}\right ] 
\quad\quad\quad\quad\quad\quad\quad\quad\quad\nonumber\\
&= -J \sum_{j} \Lambda_{ij} \left[ v_i^L\bn +\bpi_j 
%\right. \nonumber \\
- (v_i^Lv_j^L+\bpi_i\cdot\bpi_j )\frac{(v_i^L\bn+\bpi_i)}{v_0^2}\right ] \nonumber \\
&= -J \sum_{j} \Lambda_{ij} [ \bpi_j + \mathrm{O}(\bpi^2)] \ .
\end{align}
The second term at the r.h.s. of Eq.~(\ref{eqv}) is of order $\mathrm{O}(\bpi^2)$, and we therefore get 
\beq
\chi \frac{d^2 \bpi_i}{dt^2} =-J \sum_{j} \Lambda_{ij}  \bpi_j -\eta\frac{d\bpi_i}{dt}+v_0{\boldsymbol \xi}_i \times \bn \ .
\label{eqpi-a}
\eeq
The deterministic part of this expression is precisely Eq.~(\ref{eqpi}) of the main text. We notice that force contribution in the r.h.s. of (\ref{eqpi-a}) can also be obtained starting directly from the pseudo-Hamiltonian $H_{\bv}$ for the velocities. Expanding to order $\bpi^2$ we have 
\beq
H_{\bv}=\frac{J}{4v_0^2}\sum_{ij} (\bv_i-\bv_j)^2\approx \frac{J}{4v_0^2}\sum_{ij} (\bpi_i-\bpi_j)^2  \ ,
\eeq
so that $v_0^2 \bF_i^\perp \rightarrow -v_0^2 \partial H_{\bv}/\partial \bpi_i=-J \sum_j \Lambda_{ij}\bpi_j$.

\subsubsection{Low temperature expansion for ISM+FPUT}

The low temperature expansion for the ISM+FPUT proceeds very similarly to the standard ISM case analyzed above. The force appearing in Eq.(\ref{eqv}) in this case contains a non-linear term, due to the FPUT quartic term in the pseudo-Hamiltonian (\ref{pepito}), i.e.
\beq
\bF_i = - \frac{\partial H}{\partial \bv_i}
=-\frac{1}{v_0^2}\sum_{ij} n_{ij} (\bv_i-\bv_j) \left [ J + \frac{2J_4}{v_0^2}  (\bv_i-\bv_j)^2\right ] 
\label{eqv-fput}
\eeq
The expansion in the $\{\bpi_i\}$'s now has to be performed more carefully, since we want to retain terms up to the order of the new added non-linear contribution. At the level of the Hamiltonian, we then need to expand the square root of $v_i^L$ to order $\bpi^4$. One gets
\beq
(\bv_i-\bv_j)^2 \approx (\bpi_i - \bpi_j)^2 + \frac{1}{4v_0^2} (\bpi_i ^2- \bpi_j^2)^2 \ ,
\eeq
and $(\bv_i-\bv_j)^4 \approx (\bpi_i-\bpi_j)^4$. The spin-wave expansion of Eq.\eqref{pepito} is therefore given by, 
\bea
H =&& \frac{J}{4v_0^2} \sum_{ij} n_{ij} (\bv_i-\bv_j)^2 + \frac{J_4}{4v_0^4}\sum_{ij} n_{ij} (\bv_i-\bv_j)^4 \approx 
\nonumber \\
&& \frac{1}{4v_0^2}\sum_{ij} n_{ij} \left[ (\bpi_i - \bpi_j)^2 + \frac{J}{16v_0^2} (\bpi_i ^2- \bpi_j^2)^2 \right] + \nonumber \\
&& \frac{J_4}{4v_0^4} \sum_{ij} n_{ij} (\bpi_i-\bpi_j)^4   \ .
\eea
We recall now that in order to have overdamped spontaneous fluctuations, but underdamped external perturbation, we need to take $J \ll J_4$, 
which gives,
\bea
H \approx&& \frac{1}{4v_0^2}\sum_{ij} n_{ij} \left[ J  (\bpi_i - \bpi_j)^2 + \frac{J_4}{v_0^2}  (\bpi_i-\bpi_j)^4 \right] = \nonumber \\
&& \frac{1}{4v_0^2}\sum_{ij} n_{ij}  (\bpi_i - \bpi_j)^2 \left[ J  + \frac{J_4}{v_0^2}  (\bpi_i-\bpi_j)^2 \right] 
\eea
The equation for $\bpi_i$ can be quickly obtained by computing the force as the derivative of this expression and adding noise and dissipation. We then get,
\begin{align}
\chi \frac{d^2 \bpi_i}{dt^2}& =-\sum_{j} n_{ij}  (\bpi_i -\bpi_j)[J+\frac{2J_4}{v_0^2}(\bpi_i -\bpi_j)^2]\nonumber\\
& -\eta\frac{d\bpi_i}{dt}+ v_0 {\boldsymbol \xi}_i \times \bn \ .
\label{eqpi-fput}
\end{align}

To derive the semi-quantitative inequalities of the main text it is convenient to work in the simplified planar framework, namely in the case in which 
the velocities only fluctuate on a plane (this is not far from reality, incidentally, as birds tend to fly and fluctuate exclusively onto the plane orthogonal to gravity \cite{attanasi+al_14});
in this case, the transverse fluctuations become scalar variables, $\bpi_i =\pi_i$ and we can associate to each of them a dimensionless phase $\varphi_i=\pi_i/v_0$, so that, 
\beq
H\approx\frac{1}{4} \sum_{ij} n_{ij}  \, (\varphi_i-\varphi_j)^2 \left[ J +  J_4 (\varphi_i-\varphi_j)^2 \right] \ ,
\eeq
which is  expression \eqref{todi} of  the main text. The dynamical equation for $\varphi_i$ can be directly obtained from Eq.\eqref{eqpi-fput}, giving,
\begin{align}
\chi \frac{d^2 \varphi_i}{dt^2}& =-\sum_{j} n_{ij}  (\varphi_i -\varphi_j)[J+2J_4(\varphi_i -\varphi_j)^2]\nonumber\\
& -\eta\frac{d\varphi_i}{dt}+ \xi_i 
\label{eqphi-fput}
\end{align}

%%%%%%%%%%%%%%%%%%%%%%%%%%%%%%%%%%%%%%%
\subsection{Numerical simulations}

To simulate ISM and ISM+FPUT we have started from the second-order equation of motion \eqref{eqv}, where the spin variable has been removed in favour of the second derivative of the velocity.  This formulation is more convenient because it allows us to leverage the collective experience in stochastic dynamics simulations.  We have thus simulated \eqref{eqv} treating it as an inertial Langevin equation and integrated it with the procedure described in \cite{Allen1987} (sec.~9.3, eq.~9.24), which reduces to the velocity Verlet integrator for Molecular Dynamics 
\cite{Allen1987, swope_computer_1982-1} in the limit $\eta\to0$.

We rewrite \eqref{eqv} as
\begin{equation}
    \frac{d^2\mathbf{v}_i}{dt^2} = \frac{v_0^2}{\chi} \left[ \mathbf{F}_i  -\frac{\eta}{v_0^2} \frac{d \mathbf{v}_i}{dt} +\frac{\xi_i}{v_0} + \mathbf{f}_c \right]
    \label{eq:apppendix-differentialEq}
\end{equation}
where $\mathbf{F}_i = -\partial H/\partial \mathbf{v}_i$ and $\mathbf{f}_c$ represents the terms that enforce the constraint $\lvert \mathbf{v}_i\rvert = v_0$.  In \eqref{eqv} these are given by the projectors and the term containing the squared acceleration.  If one could integrate \eqref{eqv} exactly, $\mathbf{f}_c$ would guarantee a constant speed, but when time is discretized they do not correctly enforce the constrain.  To have the constraint strictly enforced one can resort to a procedure such as the RATTLE algorithm (as was done in \cite{cavagna2016spatio}), but here for simplicity we use a soft constraint, employing an harmonic restorative force $\mathbf{f}_c = -\partial V_c/\partial \mathbf{v}_i$, with $V_c = (\kappa/2) \left( \lvert\mathbf{v}_i \rvert - v_0 \right)^2$.

Then, following \cite{Allen1987}, the discretized equations are found integrating \eqref{eq:apppendix-differentialEq} in a small time interval $h$ and assuming the deterministic forces vary linearly in time.  Defining $\mathbf{a} = \frac{d\mathbf{v}}{dt}$, $\mathbf{b} = \frac{d \mathbf{a}}{dt}$ one arrives at
\begin{align}
\mathbf{v}(t+h) &= \mathbf{v}(t)+ h c_1\mathbf{a}(t)+ h^2 c_2 \mathbf{b}(t) +  \Theta_v(t) , \\
\mathbf{a}(t+h) &= c_0 \mathbf{a}(t)+(c_1-c_2) h  \mathbf{b}(t) + c_2 h  \mathbf{b}(t+h)  + \Theta_a(t), \\
  \mathbf{b}(t+h) &= \frac{v_0^2}{\chi} \left[ \mathbf{F}(\mathbf{r}(t+h))
                    + \mathbf{f_c}(\mathbf{r}(t+h)) \right],
\end{align}
where 
\begin{align}
    &c_0 = e^{ - \eta h / \chi} \\
    &c_1 = \frac{\chi }{\eta h} (1-c_0)\\
    &c_2 = \frac{\chi }{\eta h}  (1-c_1),
\end{align}
and $\Theta_v$ and $\Theta_a$ are random variables related to the random force. They are independent for each axis and each pair of components is drawn form a bivariate Gaussian distribution with zero first moments and second moments given by
\begin{align}
    \left\langle \Theta_v^2 \right\rangle &= \frac{T v_0^2}{\eta} \left[2 h -\frac{\chi}{\eta}(3-4c_0+c_0^2) \right] \\
    \left\langle \Theta_a^2 \right\rangle &= \frac{T v_0^2}{ \chi} \left( 1-c_0^2 \right) \\
    \left\langle \Theta_v \Theta_a \right\rangle  &= \frac{T v_0^2}{\eta} \left(1-c_0 \right)^2. 
\end{align}

The algorithm proceeds as in the velocity Verlet, first drawing $\Theta_v$ and $\Theta_a$ and using $\mathbf{b}(t)$ and $\mathbf{a}(t)$ to obtain $\mathbf{v}(t+h)$ and a partial update of $\mathbf{a}$.  Then $\mathbf{b}(t+h)$ is evaluated using the new velocities and the update of $\mathbf{a}$ is completed.  Finally, the positions are updated with a simple Taylor series predictor step,
\begin{equation}
\mathbf{r}(t+h) = \mathbf{r}(t)+ h \mathbf{v}(t) +(h^2/2) \mathbf{a}(t).
\end{equation}

We have set $h=10^{-2}$, $\chi=1.25$, $J=20$,  $v_0=0.1$, $T=10^{-6}$, $\kappa=10^4$. $\eta=0.7$ in the underdamped ISM simulations, while  $\eta=7$ in the overdamped ISM and in the ISM+FPUT simulations, where we have also set $J_4=10^5$ .

The correlation functions were obtained by simulating the system in a cubic box with periodic boundary conditions at a density $\rho = 1$. Perturbed simulations, on the other hand, were performed within a much larger cubic box. In this setup, the global density was $\rho = 10^{-3}$, while the system occupied only a small fraction of the total volume, maintaining a local density of $\rho \sim 1$. This latter approach was adopted for computational convenience and is effectively equivalent to a simulation with open boundary conditions.

%%%%%%%%%%%%%%%%%%%%%%%%%%%

\subsection{Out-of-equilibrium field theory of the ISM}

In the main text, we have reported the linear dispersion relation for the ISM under the approximation that the interaction network $n_{ij}$ does not depend on time (quasi-equilibrium approximation); at the coarse-grained level, this is the same as assuming that there are no density fluctuations.
This is not strictly true, of course, hence we develop here the linear theory for the coarse-grained version of the ISM, including the dynamics of density fluctuations. However, we will show in what follows that the results obtained in the quasi-equilibrium approximation remain qualitatively valid even when activity is taken into account.\par
The coarse-grained theory describing the hydrodynamic limit of the out-of-equilibrium ISM involves three fields: the velocity field $\bv(\mathbf{x},t)$, the spin field $\bs(\mathbf{x},t)$, and the density field $\rho(\mathbf{x},t)$. To ensure a well-defined equilibrium limit, one can  introduce an orientational field $\boldsymbol{\psi}$ defined as $\bv(\mathbf{x},t) = v_0 \boldsymbol{\psi}(\mathbf{x},t)$, where $v_0$ is the microscopic velocity. This approach was followed in~\cite{cavagna2021equilibrium} and~\cite{cavagna2023natural}.
However, this definition renders the field $\boldsymbol{\psi}$ dimensionless, which means that it does not have the conventional naive scaling dimension of a field, namely $d^{\text{naive}}_{\psi} = (d - 2)/2$~\cite{zinnjustin_QFTCF}. We here resolve this inconvenient feature by redefining the orientational field $\boldsymbol{\psi}$ as,
\begin{equation}
	\label{psi}
	\bv(\mathbf{x},t) = \hat{v}_0 \boldsymbol{\psi}(\mathbf{x},t)\, ,
\end{equation}
where $\hat{v}_0$ is proportional to the microscopic velocity $v_0$ through a dimensional constant. This dimensional constant is chosen in such a way that the field $\boldsymbol{\psi}$ retains the standard field dimension $d^{\text{naive}}_{\psi} = (d - 2)/2$, while ensuring that $\hat{v}_0$ vanishes when $v_0$ does. \par
The hydrodynamic theory for the ISM was first developed in~\cite{cavagna2015silent}. In that work, the theory was formulated in the planar case and longitudinal fluctuations were disregarded, as well as dissipative terms in the orientational field dynamics. Since our goal is to provide a complete treatment of the hydrodynamic limit of the ISM, we will not adopt these simplifications. 
Furthermore, subsequent studies showed that a transport term had been missing in the spin dynamics~\cite{cavagna2019renormalizationl, cavagna2019dynamical}. The hydrodynamic theory is described by the following set of equations,
\begin{subequations}
	\label{SP-MG}
	\begin{align}
		D_t \boldsymbol{\psi} & = - \Gamma \frac{\delta H}{\delta \boldsymbol{\psi}} - g \boldsymbol{\psi}\times \frac{\delta H}{\delta \mathbf{s}} - \frac{1}{\hat{v}_0}\nabla P + \boldsymbol{\zeta} \\
		D_t \mathbf{s} &= (\lambda \nabla^2 - \eta) \frac{\delta H}{\delta \mathbf{s}} - g \boldsymbol{\psi} \times \frac{\delta H}{\delta \boldsymbol{\psi}} + \boldsymbol{\theta} \\
		\partial_t \rho &= - \hat{v}_0\, \nabla \cdot (\rho \boldsymbol{\psi}) \, , \label{contEq}
	\end{align}
\end{subequations}
%where $\Gamma$ is the kinetic coefficient, $\lambda$ is the transport coefficient, 
where $\Gamma$ and $\lambda$ are the dissipative relaxation coefficients for $\psi$ and $s$ respectively, $\eta$ is the spin friction coefficient, and $g$ is the non-dissipative coupling constant. The Hamiltonian $H$ is the standard Landau-Ginzburg one, with the addition of a kinetic term for the spin field $\bs$,
\begin{equation}
	H = \frac{1}{2} \int d^{d} \mathbf{x} \left( (\nabla \boldsymbol{\psi})^2 - r\, \psi^2 + \frac{1}{2} w \, \psi^4 +  \frac{s^2}{\chi} \right)\quad ,
\end{equation}
with $r > 0$ in the ordered phase. The inertia $\chi$ will be set to 1 in the following, as its specific value does not influence the behavior of the theory.
It is important to note that the absence of a dimensional constant $\alpha^2$ in front of the $(\nabla \boldsymbol{\psi})^2$ term would not be possible if the field $\boldsymbol{\psi}$ were dimensionless. Indeed, the definition~\eqref{psi} can also be obtained by first defining the standard dimensionless orientation such that $\bv = v_0 \boldsymbol{\psi}$, and then reabsorbing the parameter $\alpha$ into it, thereby obtaining $\bv= \hat{v}_0 \boldsymbol{\psi}$, with $\hat{v}_0 = \alpha \, v_0$.\par
This theory incorporates both reversible mode-coupling cross terms, linking the orientational field $\boldsymbol{\psi}$ and the spin field $\bs$, and dissipative diagonal terms, given by $\delta_{\boldsymbol{\psi}}H$ and $\delta_{\bs}H$. These last are complemented by Gaussian noise terms, $\boldsymbol{\zeta}$ and $\boldsymbol{\theta}$, with variances,
\begin{equation}
	\label{noise}
	\begin{aligned}
		\langle \zeta_{\alpha} (\mathbf{x},t) \zeta_{\beta} (\mathbf{x}',t') \rangle &= 2 \widetilde{\Gamma} \, \delta_{\alpha \beta} \, \delta(t - t') \, \delta^{(d)}(\mathbf{x} - \mathbf{x}') \\
		\langle \theta_{\alpha} (\mathbf{x},t) \theta_{\beta} (\mathbf{x}',t') \rangle &= -2 \widetilde{\lambda} \, \nabla^2 \delta_{\alpha \beta} \, \delta(t - t') \, \delta^{(d)}(\mathbf{x} - \mathbf{x}') \quad .
	\end{aligned}
\end{equation}

The out-of-equilibrium nature of the theory is evident from several features: the presence of the pressure term $\nabla P$, which couples the orientational field $\boldsymbol{\psi}$ to the density $\rho$; the continuity equation~\eqref{contEq} and the material derivatives $D_t = \partial_t + \gamma (\hat{v}_0 \boldsymbol{\psi} \cdot \nabla)$, with $\gamma$ breaking Galilean invariance~\cite{toner+al_98}. These active terms were first introduced in~\cite{toner+al_95} within the Toner-Tu (TT) theory, which describes the hydrodynamic limit of the Vicsek Model, which is equivalent to the ISM when the spin-velocity coupling is neglected.\par
In the equilibrium limit $\hat{v}_0 \to 0$, all these terms should vanish, implying that the pressure term must be at least of order $\hat{v}_0^2$. The resulting theory describes the dynamics of $\psi$ and $s$ on a fixed network, and is given by Model G from~\cite{hohenberg1977theory} with the addition of the spin friction $\eta$. Depending on the strength of this friction, a crossover between an underdamped and overdamped dynamics is known to arise~\cite{cavagna2019renormalizationl}. This means that for sufficiently small values of $\eta$, the equilibrium theory behaves like Model G, whereas for large $\eta$ its behavior approaches that of Model A~\cite{hohenberg1977theory}. This should not come as a surprise, since from a microscopic point of view, the large friction limit of the ISM is precisely the Vicsek Model, whose equilibrium counterpart is described by Model A. Note that a similar crossover, between an active version of Model G and the Toner and Tu theory, also arises in the critical non-equilibrium case~\cite{cavagna2023natural}.\par
We shall focus our attention, from now on, to the fully polarized phase. Here, the field $\boldsymbol{\psi}$ spontaneously breaks the rotational symmetry, and can thus be expanded around its Landau mean-field solution $\psi_0 = \sqrt{r/w}$, as follows~\cite{ryder1996quantum},
\begin{equation}
	\label{swexp}
	\boldsymbol{\psi} = \psi_0 \hat{x} + \boldsymbol{\varphi} \, ,
\end{equation}
where the fluctuations satisfy $\varphi_\alpha \ll \psi_0$, and we assume the polarization is aligned along the $x$-axis.
Substituting expression~\eqref{swexp} into the functional derivative of the Hamiltonian and retaining only linear terms, we obtain,
\begin{equation}
	\begin{aligned}
		-\frac{\delta H}{\delta \boldsymbol{\psi}} &= \nabla^2 \boldsymbol{\psi} + (r - w \psi^2) \boldsymbol{\psi} \\
		&= \nabla^2 (\psi_0 \hat{x} + \boldsymbol{\varphi}) + \left(r - w (\psi_0 \hat{x} + \boldsymbol{\varphi})^2\right)(\psi_0 \hat{x} + \boldsymbol{\varphi}) \\
		&\sim \nabla^2 \boldsymbol{\varphi} - 2r \varphi_x \hat{x} \quad + \text{non-linearities}\, .
	\end{aligned}
\end{equation}
Moreover, it is possible to linearize the density and the pressure terms by writing the first as $\rho = \rho_0 + \delta \rho$, and the second as $P = \sigma \hat{v}_0^2 \, \delta \rho$. We remind that the factor $\hat{v}_0^2$ in the pressure term ensures that the equilibrium limit is correctly recovered by taking $\hat{v}_0 \to 0$, with $\sigma$ remaining finite in this limit.
Under the linear approximation, the material derivative becomes
\begin{equation}
	D_t \simeq \partial_t + \gamma \hat{v}_0 \psi_0 \, \partial_x \, .
\end{equation}
The term $\gamma \hat{v}_0 \psi_0 \, \partial_x$ can be eliminated by the following change of variables,
\begin{equation}
	\begin{aligned}
		t' &= t \\
		\mathbf{x}' &= \mathbf{x} + \gamma \, v_0 \, \psi_0 \, t \, \hat{x} \, .
	\end{aligned}
\end{equation}
With this transformation, in the linear theory, the material derivative simplifies to a partial derivative with respect to time. Note that this transformation is equivalent to switching to the center-of-mass reference frame, which is what was performed in experimental and numerical analyses, as long as $\gamma$ is close to $1$. Finally, we can further simplify the analysis by exploiting the rotational symmetry of the system in the $yz$-plane. As a result, we can, without loss of generality, consider wavevectors $\mathbf{k}$ satisfying $k_z = 0$.
After all these steps, the resulting linearized equations in Fourier space are,
\begin{equation}
	\begin{aligned}
		-i \omega \varphi_x &= -\Gamma k^2 \varphi_x - 2 \Gamma r \varphi_x - i \hat{v}_0 \hat{\sigma} k_x \delta \rho + \zeta_x\\
		-i \omega \varphi_y &= -\Gamma k^2 \varphi_y - g \psi_0 s_z - i \hat{v}_0 \hat{\sigma} k_y \delta \rho + \zeta_y \\
		-i \omega \varphi_z &= -\Gamma k^2 \varphi_z + g \psi_0 s_y + \zeta_z \\
		-i \omega s_x &= -(\lambda k^2 + \eta) s_x + \theta_x\\
		-i \omega s_y &= -(\lambda k^2 + \eta) s_y - g \psi_0 k^2 \varphi_z + \theta_y\\
		-i \omega s_z &= -(\lambda k^2 + \eta) s_z + g \psi_0 k^2 \varphi_y + \theta_z\\
		-i \omega \delta \rho &= -i \hat{v}_0 \rho_0 (\mathbf{k} \cdot \boldsymbol{\varphi}) 
	\end{aligned}
\end{equation}

where the $x$-component of the spin field can be omitted because it exhibits purely relaxational dynamics, decoupled from the other fields. The equations for $\varphi_z$ and $s_y$ are also decoupled from the rest and do not depend on $\hat{v}_0$, meaning that $\varphi_z$ and $s_y$ evolve according to the equilibrium dynamics.\par

We aim to compute the connected correlation functions of the velocity field,
\begin{equation}
	\begin{aligned}
		C_\alpha(\mathbf{k}, \omega) &= \langle v_\alpha(-\mathbf{k}, -\omega) v_\alpha(\mathbf{k}, \omega) \rangle_C \\
		&= \hat{v}_0^2 \; \langle \varphi_\alpha(-\mathbf{k}, -\omega) \varphi_\alpha(\mathbf{k}, \omega) \rangle \, ,
	\end{aligned}
\end{equation}
where the subscript $C$ denotes the connected part of the correlation function. 
From $C_\alpha$, the density correlation function can be readily obtained using the continuity equation. In particular, when the wavevector $\mathbf{k}$ is aligned along the coordinate axis $\alpha$, namely $\mathbf{k}=k\boldsymbol{\hat{\alpha}}$, the connected correlation function of the density fluctuations is given by,
\begin{equation}
	S(k \boldsymbol{\hat{\alpha}}, \omega) = \langle \delta \rho(-\mathbf{k}, -\omega) \, \delta \rho(\mathbf{k}, \omega) \rangle = \frac{k^2}{\omega^2} \, C_\alpha(k \boldsymbol{\hat{\alpha}}, \omega) \,.
\end{equation}
\\

\begin{widetext}
	The correlation function $C_z(\mathbf{k},\omega)$ is obtained from the dynamics of $\varphi_z$ and $s_y$, and reads,
	\begin{equation}
		\begin{aligned}  
			\label{cz}
			&C_z(\mathbf{k},\omega)= \frac{
				2 \hat{v}_0^2\left(c_2^4 k^2 \widetilde{\lambda} + \widetilde{\Gamma} \left((\eta + k^2 \lambda)^2 + \omega^2\right)\right)}{
				k^4 \left(c_2^4 + \Gamma (\eta + k^2 \lambda)\right)^2 + \left(\eta^2 + k^2 \left(-2 c_2^4 + 2 \eta \lambda + k^2 \left(\Gamma^2 + \lambda^2\right)\right)\right) \omega^2 + \omega^4
			}\,.
		\end{aligned}
	\end{equation}
	\par
	From the linearized equations for $\varphi_x$, $\varphi_y$, $s_y$, and $\delta \rho$, it is possible to solve the linear system, obtaining, 
	
	\begin{align}
		C_x(\mathbf{k},\omega) 
		&= \frac{2 \hat{v}_0^2}{|\Delta (\mathbf{k},\omega)|^2} \Bigg[ 
		k_x^2 k_y^2 \hat{v}_0^4 c_1^4 \Big( 
		c_2^4 k^2 \widetilde{\lambda}
		+ \widetilde{\Gamma} \left[ (\eta + k^2 \lambda)^2 + \omega^2 \right]
		\Big) \notag \\
		&\quad + \widetilde{\Gamma} \Big(
		k^4 \left( c_2^4 + \Gamma \eta + k^2 \Gamma \lambda \right)^2 \omega^2 
		+ \left[ \eta^2 + k^2 \left( -2 c_2^4 + 2 \eta \lambda + k^2 (\Gamma^2 + \lambda^2) \right) \right] \omega^4 
		+ \omega^6 \notag \\
		&\quad + k_y^4 \hat{v}_0^4 c_1^4 \left[ (\eta + k^2 \lambda)^2 + \omega^2 \right]
		- 2 k_y^2 \hat{v}_0^2 c_1^2 \omega^2 \left( -c_2^4 k^2 + (\eta + k^2 \lambda)^2 + \omega^2 \right)
		\Big)
		\Bigg],
	\end{align}
	
	and
	
	\begin{align}
		C_y(\mathbf{k},\omega) 
		&= \frac{2 \hat{v}_0^2}{|\Delta (\mathbf{k},\omega)|^2} \Bigg[
		k_x^4 \hat{v}_0^4 c_1^4 \Big(
		c_2^4 k^2 \widetilde{\lambda} 
		+ \widetilde{\Gamma} \left[ (\eta + k^2 \lambda)^2 + \omega^2 \right]
		\Big)
		+ \omega^2 \left[ (k^2 + 2 r)^2 \Gamma^2 + \omega^2 \right] \notag \\
		&\quad \times \Big(
		c_2^4 k^2 \widetilde{\lambda} 
		+ \widetilde{\Gamma} \left[ (\eta + k^2 \lambda)^2 + \omega^2 \right]
		\Big)
		+ k_x^2 \hat{v}_0^2 c_1^2 \bigg(
		k_y^2 \hat{v}_0^2 \widetilde{\Gamma} c_1^2 \left[ (\eta + k^2 \lambda)^2 + \omega^2 \right] \notag \\
		&\quad - 2 \omega^2 \left(
		c_2^4 k^2 \widetilde{\lambda} 
		+ \widetilde{\Gamma} \left[ (\eta + k^2 \lambda)^2 + \omega^2 \right]
		\right)
		\bigg)
		\Bigg],
	\end{align}
	where $k^2=k_x^2+k_y^2$ and where $\Delta$ is the determinant of the linear system,
	\begin{align}
		\Delta(\mathbf{k},\omega) 
		&= c_2^{4} k^{2} \Big( k_x^{2} \hat{v}_0^{2} c_1^2 
		- i \left( k^{2} \Gamma + 2 r \Gamma - i \omega \right) \omega \Big) 
		+ \left( \eta + k^{2} \lambda - i \omega \right) \notag\\
		& 
		\quad \times \Big( k_y^{2} \hat{v}_0^{2} c_1^2 \left( 2 r \Gamma - i \omega \right) 
		- i k^{4} \Gamma^{2} \omega 
		+ k^{2} \Gamma \left( k^2 \hat{v}_0^{2} c_1^2  - 2 i r \Gamma \omega - 2 \omega^{2} \right) \notag\\
		&\quad
		+ i \omega \left( - k_x^{2} \hat{v}_0^{2} c_1^2 + 2 i r \Gamma \omega + \omega^{2} \right) \Big)
		\, .
	\end{align}
\end{widetext}
We have defined the velocity of first sound as $c_1^2 = \sigma \rho_0$ and of second sound as $c_2^2 = g \psi_0$, in analogy with~\cite{cavagna2015silent}. Notice that, having defined the pressure proportional to $\hat{v}_0^2$, the actual first sound velocity is $\hat{v}_0 c_1$, but we chose not to absorb the factor $\hat{v}_0$ into the definition to keep it explicitly present in the equations.
First sound is the wave-like propagation of orientational fluctuations arising from the density-velocity coupling~\cite{toner+al_98}, and hence is intrinsically related to the presence of density waves. In contrast, second sound corresponds to the spin-mediated propagation of the orientation field and occurs even when the density field is held fixed, i.e., at equilibrium.

\begin{figure}
	\centering
	\includegraphics[width=\linewidth]{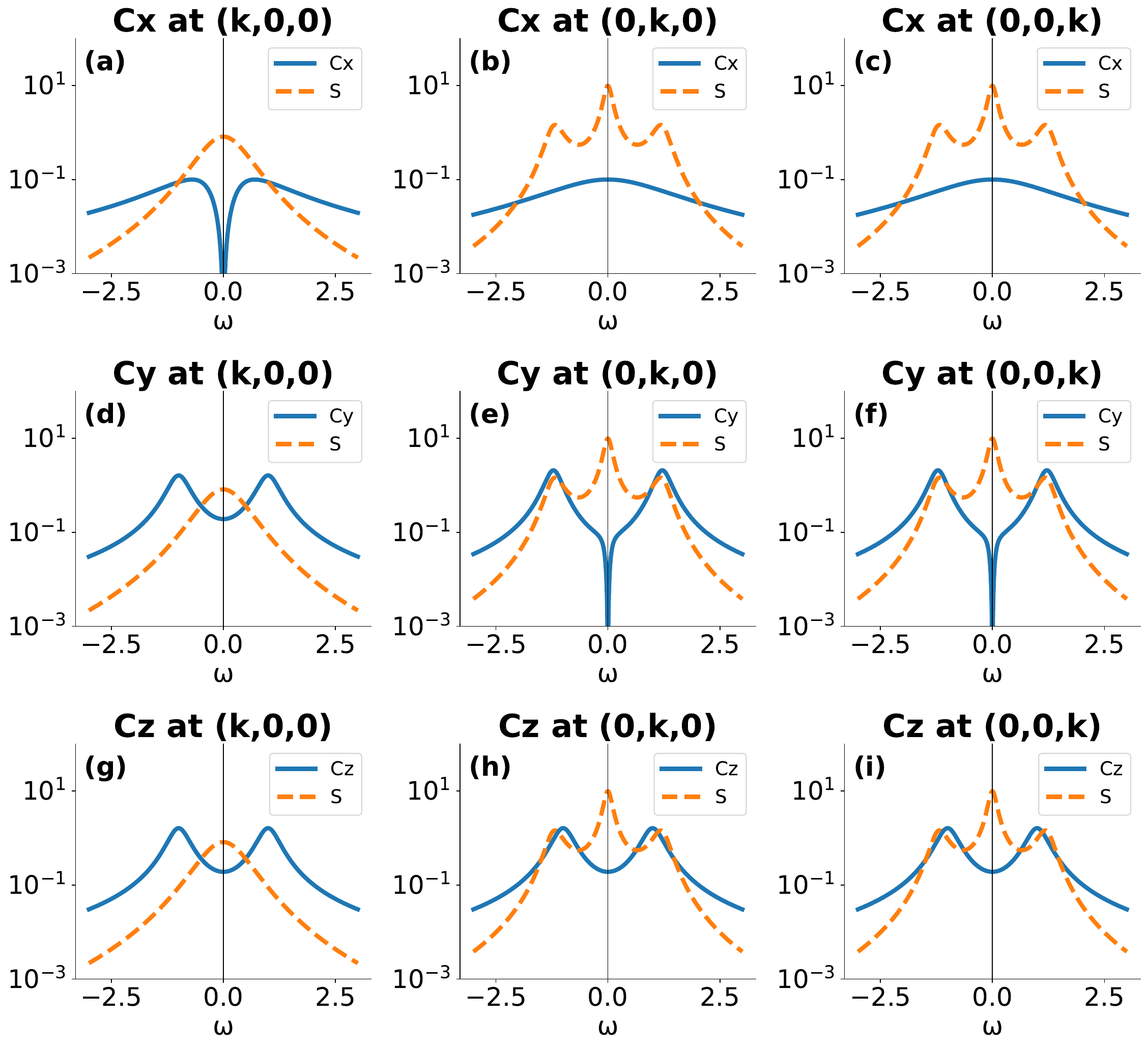} 
	\caption{The correlation functions $C_x$ (panels (a), (b), and (c)), $C_y$ (panels (d), (e), and (f)), and $C_z$ (panels (g), (h), and (i)) as functions of $\omega$, together with $S(k,\omega)$. The correlation functions are displayed in the underdamped regime for $\mathbf{k} = (k,0,0)$, $\mathbf{k} = (0,k,0)$, and $\mathbf{k} = (0,0,k)$. In panels (a), (e), and (i), it is clear how the continuity equation forces $C_\alpha$ to vanish when $\mathbf{k}$ is aligned with the $\alpha$-axis. Apart from that, $C_y$ and $C_z$ display the usual spin-wave correlation function, while $C_x$ has a Lorentzian correlation function, with values much smaller than those of $C_y$ and $C_z$. The $y$-axis is shown on a logarithmic scale. Here $\hat{v}_0 = 0.7$, $k=1$, $c_1 = c_2 = 1$, $\Gamma = \widetilde{\Gamma} = 0.2$, $\lambda = \widetilde{\lambda} = 0.2$, $\eta = 0.1$ and $r = 3$.}
	\label{FIG7}
\end{figure}

%\textcolor{orange}{Commento per me stesso: Vanno collegate meglio questa sezione sui poli e quella dopo sull'effettiva forma di C.}  %Let us start our analysis by looking at the characteristic frequencies of the different modes, given by the poles of the corresponding correlation function.
The characteristic frequencies of the different modes can give important insights into how information propagates throughout the flock. Furthermore, we shall see how these frequencies are closely related to the shape of the correlation functions, measurable in experiments.
As previously mentioned, $\varphi_z$ -- namely the velocity mode orthogonal to both the flocking direction $\hat{x}$ and the wavevector $\boldsymbol{k}$ -- and its conjugated spin in the $yz$ plane $s_y$ decouple from the rest of the modes at the linear level. Because of this, the correlation functions of $\varphi_z$ and $s_y$ are virtually identical to those of the equilibrium case, with characteristic frequencies given by
\begin{equation}
	\omega_z=-i\frac{\eta + k^2 (\Gamma +\lambda )}{2} \pm\sqrt{c_2^2 k^2-\frac{\left(\eta +k^2 (\lambda -\Gamma )\right)^2}{4}}
\end{equation}
It is possible to see that two main scales emerge: a friction scale $k_\eta=\eta/c_2$ and a dissipative scale $k_d=c_2/|\Gamma-\lambda|$. Whenever second-sound propagation is not too small, $k_\eta\ll k_d$, three different regimes emerge. At large scales, $k<k_\eta$, spin friction kicks in, suppressing the spin-velocity coupling. The $\varphi_z$ is thus purely diffusive with $\omega_z\sim-i\,\Gamma k^2$. On the other hand, at very small scales $k>k_d$, dissipative effects also lead to diffusive behavior. The interesting physics emerges within the interval $k_\eta<k< k_d$, where the characteristic frequency acquires a real part given by
\begin{equation}
	\omega_z \simeq \pm c_2 k\, ,
\end{equation}
a hallmark of second sound wave propagation.
For the other modes, $\varphi_x,\varphi_y,s_z,\delta\rho$, the characteristic frequency is given by the zeros of $\Delta(k,\omega)$. Deep in the ordered phase, namely when $k^2\ll r$, the mode $\varphi_x$ relaxes on the fast timescale $(\Gamma r)^{-1}$ and hence at the linear level effectively decouples from $\varphi_y$ and $\delta\rho$.\footnote{We know from \cite{pata1973chi_longi} that non-linear effects {\it enslave} $\varphi_x$ to $\varphi_y^2$, but this should not affect our current discussion on the shape of the correlation function.} 
In this limit, the general functional dependence of the characteristic frequency $\omega_y$ on $k$ is still rather cumbersome, and perhaps not very insightful. However, the overall features are not so far from those of $\omega_z$, with a few crucial differences. At large scales, $k<k_\eta$, the effects of the spin-velocity coupling are still negligible, but wave-like propagation still arises in the form of first sound $\omega\sim\pm\hat{v}_0 c_1 k_y$, which are extremely anisotropic. Second sound propagation instead emerges over intermediate scales $k_\eta<k< k_d$. Here, the characteristic frequency retains a reminiscence of the coupling with the density, leading to a modification of the second-sound dispersion relation, which now becomes~\cite{cavagna2015silent}
\begin{equation}
	\omega \simeq \pm \sqrt{\hat{v}_0^2 c_1^2 k_y^2 + c_2^2 k^2 }\, .
\end{equation}
When dissipative effects and friction are small, the imaginary part of $\omega$ is small, and the poles are close to the real axis. The corresponding correlation functions thus develop peaks in correspondence with the real part of these frequencies, leading to the characteristic double-peak spin-wave form.
The rather cumbersome correlation functions obtained, along with the density correlation function $S(\mathbf{k}, \omega)$, are shown in Figure ~\ref{FIG7} in the underdamped second sound regime and for three different wavevectors: $\mathbf{k} = (k, 0, 0)$, $\mathbf{k} = (0, k, 0)$, and $\mathbf{k} = (0, 0, k)$. It is worth noting that the case $\mathbf{k} = (0, 0, k)$ can be directly obtained from the $k_z=0$ case we just treated through the following symmetry relations:
\begin{equation}
	\begin{aligned}
		C_x((0, 0, k)) &= C_x((0, k, 0))\\
		C_y((0, 0, k)) &= C_z((0, k, 0)) \\
		C_z((0, 0, k)) &= C_y((0, k, 0)) \, ,
	\end{aligned}
\end{equation}
which follow from the rotational symmetry in the $yz$-plane.\par

In Figure~\ref{FIG7}, we observe that the continuity equation enforces $C_\alpha(\boldsymbol{k}_\alpha)$ -- the correlation functions plotted along the diagonal of Figure ~\ref{FIG7}, namely panels (a),(e) and (i) -- to vanish at $\omega = 0$. 
Aside from this constraint, the correlation functions exhibit the same qualitative behavior as in equilibrium: the transverse components (panels (f) and (h)) display spin-wave peaks at $\omega \sim \pm c_2 k$, in complete analogy with the peaks at $\omega \sim \pm c_s k$ discussed in the main text for the microscopic ISM. On the other hand, the longitudinal fluctuation exhibits purely relaxational dynamics, as evidenced by the Lorentzian shape of $C_x$. Moreover, $C_x$ has smaller values than the transverse correlation function, since fluctuations along the symmetry-breaking direction $\hat{x}$ are much smaller than those orthogonal to it in the low-temperature phase~\cite{patashinskii_book}.
As noted previously, in $C_y((0,k,0))$ and $C_z((0,0,k))$ the coupling with the density leads to a slight shift in the spin wave peak position to $\omega \sim \pm \sqrt{c_2^2 + \hat{v}_0^2 c_1^2} \; k$.
Despite this shift, there is no qualitative difference between the spin-wave peaks of $C_y((0,k,0))$ and $C_y((0,0,k))$.
In the overdamped regime, where second sound is absent, first sound still persists and leads to highly anisotropic propagation of fluctuations, since fluctuations along a given axis propagate only in that direction~\cite{cavagna2015silent}.
Interestingly, the density correlation function $S(\mathbf{k},\omega)$ also exhibits spin-wave peaks when $\mathbf{k}$ is orthogonal to the direction of motion ($x$-axis), due to its coupling with the orientational field. This is a genuinely out-of-equilibrium feature: in the equilibrium limit, the orientational field retains its spin-wave peaks, whereas the connected density correlation function vanishes.\par
\begin{figure}
	\centering
	\includegraphics[width=0.9\linewidth]{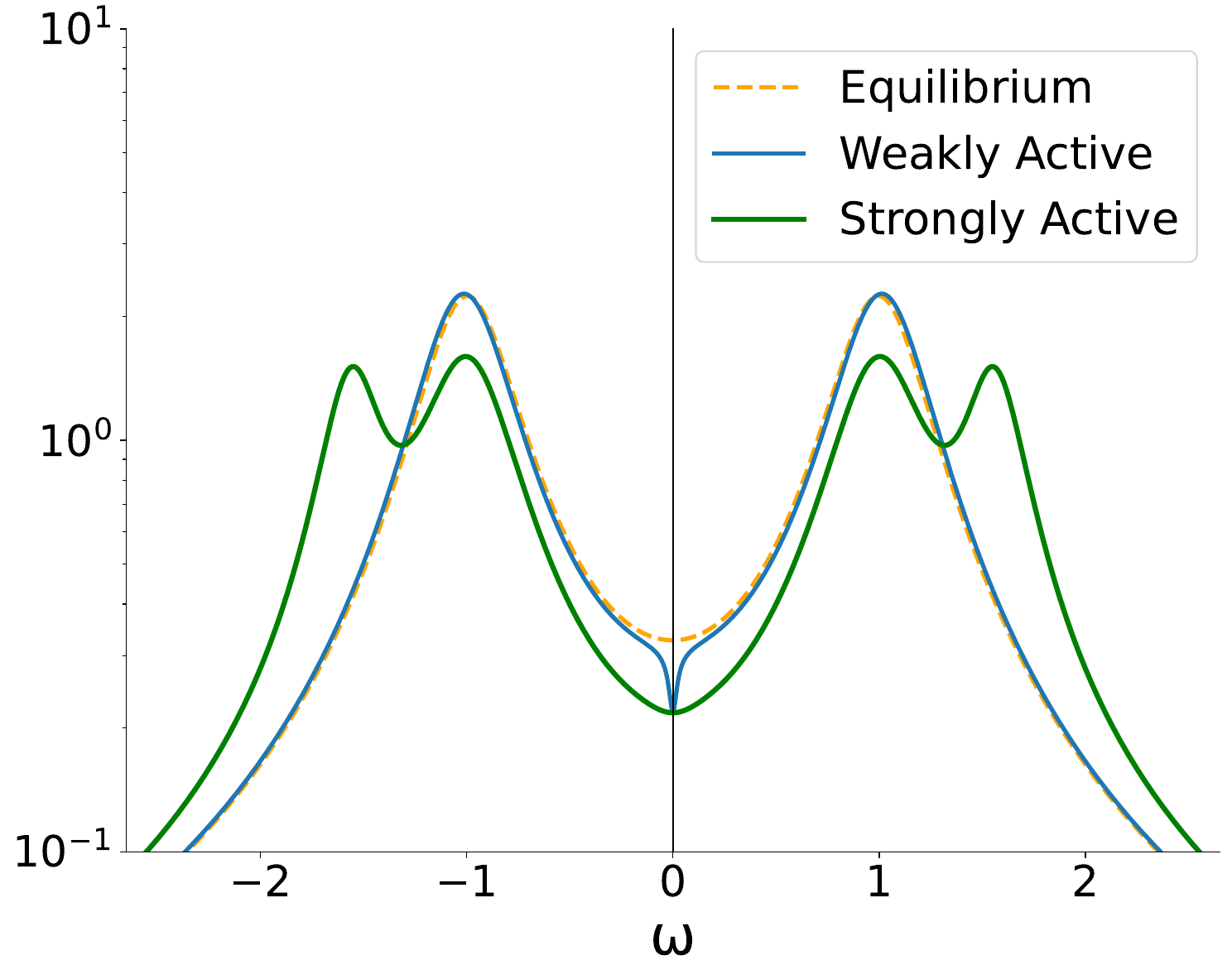} 
	\caption{The isotropic out-of-equilibrium correlation function $C_{\mathrm{ISO}}(k,\omega)$ is compared with the equilibrium correlation function of the orientational field in an underdamped regime. The equilibrium correlation function can be obtained dividing $C_{\mathrm{ISO}}(k,\omega)$ by $\hat{v}_0^2$ and taking the limit $\hat{v}_0 \to 0$. Both a weakly active case ($\hat{v}_0=0.3)$ and a strongly active case ($\hat{v}_0=1.2)$ are displayed. The active $C_{\mathrm{ISO}}(k,\omega)$ are rescaled by a factor $\hat{v}_0^{-2}$ to facilitate comparison with the equilibrium case. All three correlation functions are peaked at $\omega=\pm c_2 \,k$, while only in the strongly active case other two peaks at $\omega \sim \pm \sqrt{\hat{v}_0^2 c_1^2 + c_2^2}\;k$ appear. The $y$-axis is shown on a logarithmic scale. Here $k=1$, $c_1 = c_2 = 1$, $\Gamma = \widetilde{\Gamma} = 0.2$, $\lambda = \widetilde{\lambda} = 0.2$, $\eta = 0.1$ and $r = 3$.
	} 
	\label{FIG8}
\end{figure}
Since both experimental and numerical analyses compute the isotropic velocity correlation function, the relevant quantity to compare with simulations and experiments is given by:
\begin{equation}
	\begin{aligned}
		&C(k,\omega) = \left\langle \bv(-\mathbf{k}, -\omega) \cdot \bv(\mathbf{k}, \omega) \right\rangle \\
		&\;= C_{x,\mathrm{ISO}}(k,\omega) + C_{y,\mathrm{ISO}}(k,\omega) + C_{z,\mathrm{ISO}}(k,\omega) \\
		&\;= \frac{1}{3} \Big( 
		C_x\big((k,0,0),\omega\big) + C_x\big((0,k,0),\omega\big) + C_x\big((0,0,k),\omega\big) \\
		&\quad + C_y\big((k,0,0),\omega\big) + C_y\big((0,k,0),\omega\big) + C_y\big((0,0,k),\omega\big) \\
		&\quad + C_z\big((k,0,0),\omega\big) + C_z\big((0,k,0),\omega\big) + C_z\big((0,0,k),\omega\big) 
		\Big) \, ,
	\end{aligned}
\end{equation}
where the directionally averaged components $C_{\alpha,\mathrm{ISO}}(k,\omega)$ are obtained by averaging over the three Cartesian directions, in analogy with the procedure used in numerical simulations.
The resulting isotropic correlation function is shown in Figure~\ref{FIG8} for both the weakly active and strongly active cases, along with the corresponding isotropic correlation function of the orientation field in the equilibrium limit. In the weakly active case, the correlation function is nearly identical to the equilibrium one, differing only in the very low-frequency regime, while the overall spin-wave structure remains essentially unchanged. In this small $\hat{v}_0$ regime, the range of $\omega$ over which the equilibrium and out-of-equilibrium correlation functions differ is very narrow, making such differences almost unobservable in experiments and simulations. As the value of $\hat{v}_0$ increases, the first sound peak becomes clearly visible, and four peaks appear: two at $\omega \sim \pm c_2 k$ and two at $\omega \sim \pm \sqrt{\hat{v}_0^2 c_1^2 + c_2^2}\;k$. This behavior is observed in the strongly active case shown in Figure~\ref{FIG8} and does not seem appropriate to describe the quasi-equilibrium state of natural flocks~\cite{mora2016local}. In all cases, the isotropic correlation function is never peaked around $\omega \sim 0$ and always retains its spin-wave structure.

\end{document}